\documentclass[onecolumn,authoryear]{els-mrw} 

\usepackage{amsmath,amssymb,amsfonts,amsthm,makeidx,graphicx}
\usepackage{txfonts}
\usepackage{helvet}

\def\aj{AJ}%
\def\araa{ARA\&A}%
\def\apj{ApJ}%
\def\apjl{ApJ}%
\def\apjs{ApJS}%
\def\aap{A\&A}%
\def\mnras{MNRAS}%
\def\nar{New A Rev.}%
\def\pasp{PASP}%
\def\pasj{PASJ}%
\def\nat{Nature}%
\begin{document}

\chapter{Unification models of Active Galactic Nuclei}\label{chap1}

\author[1,2,3]{Claudio Ricci}%
%\author[2]{Second Author}%

%\author[1,2]{Third Author}%
\address[1]{Department of Astronomy, University of Geneva, ch. d'Ecogia 16, 1290, Versoix, Switzerland}
\address[2]{Instituto de Estudios Astrof\'isicos, Facultad de Ingenier\'ia y Ciencias, Universidad Diego Portales, Av. Ej\'ercito Libertador 441, Santiago, Chile}
\address[3]{Kavli Institute for Astronomy and Astrophysics, Peking University, Beijing 100871, China}

\articletag{Chapter Article tagline: update of previous edition,, reprint..}

\maketitle

\begin{glossary}[Glossary]
\term{Accretion Disk} a disk of gas fuelling the accreting black hole, where gravitational energy is converted into radiation.\newline
\term{Active Galactic Nucleus (AGN)} the central region of a galaxy, powered by accretion onto a supermassive black hole, emitting intense radiation across the whole electromagnetic spectrum.\newline
\term{Broad-Line Region (BLR)} a region close to the SMBH characterized by high-velocity gas producing broad spectral emission lines typically observed in the optical and ultraviolet.\newline
\term{Changing-Look AGN} AGN that transition between Type\,1 (unobscured) and Type\,2 (obscured) classifications over observable timescales.\newline
\term{Changing-State AGN (CS-AGN)} AGN exhibiting transitions between Type\,1 and Type\,2 states in the optical/UV, linked to variations in accretion rate.\newline
\term{Changing-Obscuration AGN (CO-AGN)} AGN where variability in obscuration along the line of sight causes apparent transitions between obscured and unobscured classifications in the X-rays.\newline
\term{Compton-thick AGN} AGN obscured by gas with column densities $N_{\rm H} \gtrsim 10^{24}\,\mathrm{cm}^{-2}$, blocking most X-rays.\newline
\term{Eddington Ratio ($\lambda_{\rm Edd}$)} the ratio of the AGN luminosity to the Eddington luminosity.\newline
\term{Hot Dust Obscured Galaxies (Hot DOGs)} very luminous AGNs with significant dust obscuration, characterized by strong infrared emission.\newline
\term{Narrow-Line Region (NLR)} a region of slow-moving gas, responsible for narrow spectral emission lines.\newline
\term{Polar Dust} dust located above and below the plane of the torus, possibly associated with AGN outflows.\newline
\term{Red Quasars} quasars exhibiting reddened optical spectra due to substantial dust obscuration.\newline
\term{Supermassive Black Hole (SMBH)} a black hole with a mass ranging from millions to billions of solar masses, typically found at the center of galaxies.\newline
\term{Torus} a dusty, anisotropic structure surrounding the AGN, obscuring or revealing the nucleus depending on the viewing angle.\newline

\end{glossary}

\begin{glossary}[Nomenclature]
\begin{tabular}{@{}lp{34pc}@{}}
AGN  & Active Galactic Nucleus \\
BLR  & Broad-Line Region \\
CL AGN& Changing-look AGN \\
CO AGN& Changing-obscuration AGN \\
CS AGN& Changing-state AGN \\
CT AGN & Compton-thick AGN \\
$f_{\rm obs}$ & Fraction of obscured AGN \\
Hot DOGs & Hot Dust Obscured Galaxies \\
IR   & Infrared \\
ISM  & Interstellar Medium \\
$L_{\rm Bol}$ & Bolometric luminosity \\
$M_{\rm BH}$ & Black hole mass, typically expressed in solar masses $M_{\odot}$ \\
$N_{\rm H}$ & Column density of intervening gas, measured in $\mathrm{cm}^{-2}$ \\
NLR  & Narrow-Line Region \\
SED  & Spectral Energy Distribution \\
SMBH & Supermassive Black Hole \\
ULIRGs & Ultra-Luminous Infrared Galaxies \\
UV & Ultraviolet \\
$\lambda_{\rm Edd}$ & Eddington Ratio \\
\end{tabular}
\end{glossary}
\clearpage

\begin{abstract}[Abstract]
This chapter presents an overview of the unification models for Active Galactic Nuclei (AGN), focusing on the physical structures, classification schemes, and evolutionary processes that characterize accreting supermassive black holes. We introduce the fundamental components of AGN, including the supermassive black hole, accretion disk, jets, outflows, broad-line and narrow-line regions, polar dust and the dusty anisotropic obscurer. The traditional orientation-based unification model is reviewed, with a focus on the role of the covering factor of the obscuring material in shaping observed properties. We introduce the radiation-regulated unification model, which accounts for the influence of radiative feedback on the nuclear environment of SMBHs. We also examine the evolutionary aspects of AGN, including the impact of galaxy mergers, host galaxy properties, and redshift-dependent trends. Finally, we address the phenomena of changing-look AGN, which challenge conventional unification frameworks by exhibiting dramatic temporal variability.
\end{abstract}

\section{Objectives}\label{sec:background}

\begin{itemize}
\item Understand the key components of AGN, including the supermassive black hole, accretion disk, and surrounding structures.
\item Explore the different classes of jetted and non-jetted AGN.
\item Learn about the traditional orientation-based unification models and how AGN classification is influenced by the viewing angle.
\item Understand the importance of the covering factor of the obscuring material and its role in unification models.
\item Explore the effect of radiation pressure on the unification model.
\item Investigate the evolutionary aspects of AGN, including their connection to galaxy mergers, host galaxy properties, and redshift-dependent trends.
\item Examine the phenomena of changing-look AGN and their implications for the current unification frameworks.
\end{itemize}

\section{Introduction}\label{sec:background}

Supermassive black holes (SMBHs), with masses typically ranging between $10^6 M_\odot$ and $10^{10} M_\odot$, are known to reside at the cores of most massive galaxies \citep[e.g.,][]{Kormendy:1995fk}. These black holes are thought to accumulate the majority of their mass through the accretion of surrounding material from their nuclear regions \citep[e.g.,][]{Soltan:1982vc}. During phases of rapid accretion, SMBHs can release vast amounts of energy across the electromagnetic spectrum, often surpassing the brightness of their host galaxies \citep[e.g.,][]{Elvis:1994wl}, and are observed as active galactic nuclei (AGNs).  These AGNs are the most luminous persistent sources of radiation in the Universe, and they are thought to play a significant role in galaxy evolution. This has been suggested by the discovery of correlations between the mass of SMBHs and key properties of their host galaxies, such as the bulge's luminosity and mass \citep[e.g.,][]{Marconi:2003ov}, as well as stellar velocity dispersion \citep[e.g.,][]{Ferrarese:2000uq, Gebhardt:2000kx,Kormendy:2013fk}. This influence is often linked to feedback processes in which energy and radiation emitted by the AGN interact with the host galaxy's interstellar medium (ISM), potentially affecting star formation rates \citep[e.g.,][]{Fabian:2012mk}. The importance of AGN feedback is supported by both semianalytical galaxy formation models \citep[e.g.,][]{Croton:2006xo} and hydrodynamic simulations \citep[e.g.,][]{Sijacki:2007xh, Schaye:2015nb}, which demonstrate its role in regulating star formation and explaining the observed galaxy mass function at the high-mass end. This feedback can manifest as either radiative feedback, commonly associated with high-luminosity AGNs \citep[e.g.,][]{Fabian:2012mk}, or kinetic feedback, often linked to AGNs of lower luminosity \citep[e.g.,][]{Weinberger:2017te}.

Over the past several decades a large number of AGN have been discovered, showing a wide array of different observational properties. While all these sources are powered by accreting SMBHs, they can display extremely different characteristics at different wavelengths. The goal of this review is to provide a summary of the current status of AGN unification models by exploring its foundations, identifying limitations and challenges, and considering recent advancements in AGN research. \S\ref{sect:componennts} describes the primary physical components of AGN, while \S\ref{sect:AGNclassification} reviews the traditional angle-dependent unification model, discussing how orientation influences AGN classification. \S\ref{sect:RRunification} introduces the radiation-regulated unification model, highlighting the critical role of the Eddington ratio in shaping the properties of the obscuring material. \S\ref{sect:evolutionMergRed} focuses on the evolutionary aspects of AGN, including the impact of galaxy mergers and redshift-dependent trends, while in \S\ref{sect:changinglook} we briefly review the phenomena of changing-look AGN.

\section{The main components of Active Galactic Nuclei}\label{sect:componennts}
\begin{figure*}
\centering
 %% 1st image
 %% 2nd image
\includegraphics[width=0.78\textwidth]{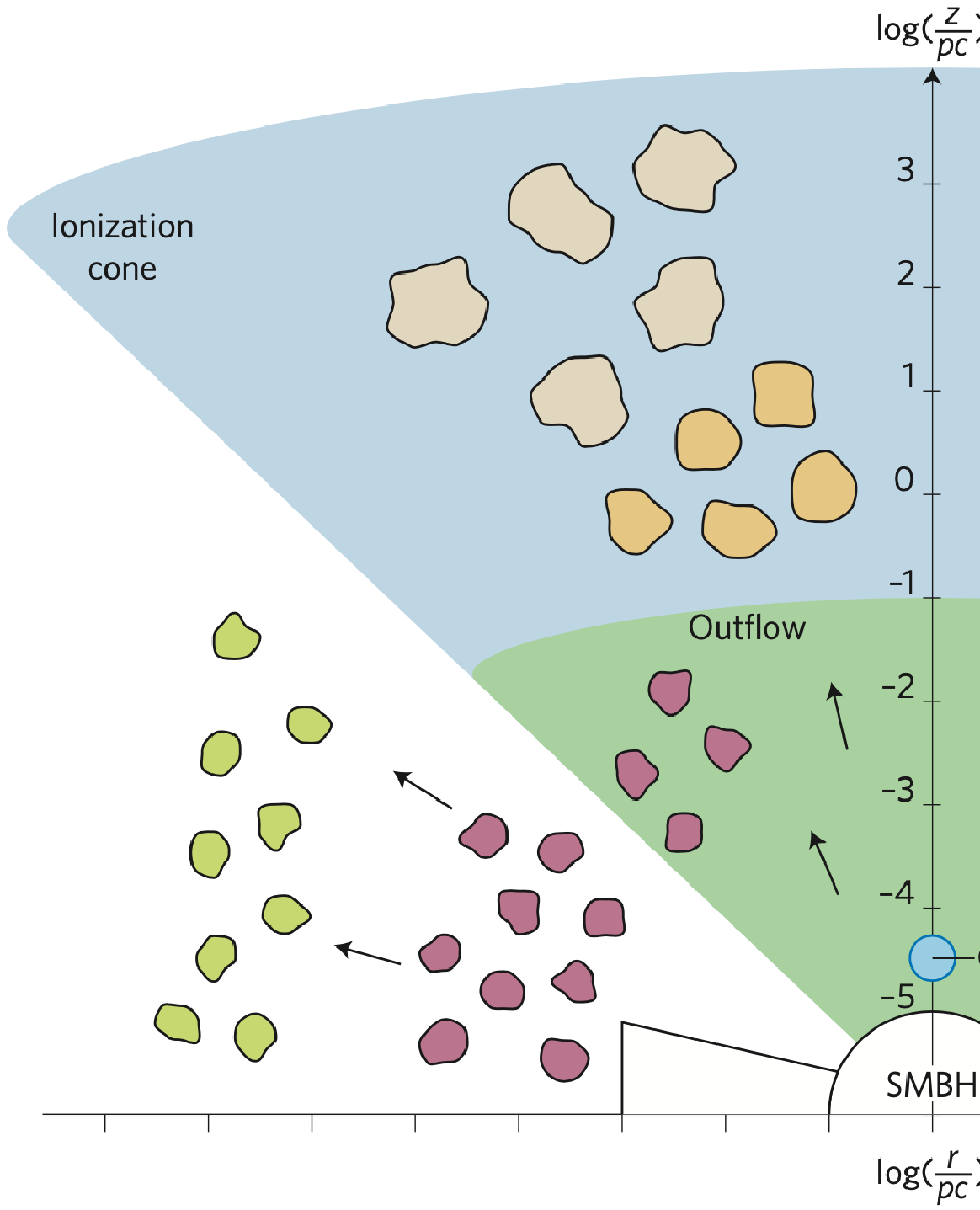}
% %% caption
  \caption{Schematic view of the the innermost regions of (non-jetted) accreting supermassive black holes. The axis show the typical scales expected in nearby AGN. Figure from \citet{Ramos-Almeida:2017ar}.}
\label{fig:UM_RR}
\end{figure*}

The current view of AGN unification suggests that all AGNs share a similar structure, which can include, from the very center to host-galaxy scales: a massive black hole (\S\ref{sect:SMBH}), a X-ray corona (\S\ref{sect:corona}), an accretion disk (\S\ref{sect:disk}), a relativistic jet (\S\ref{sect:jets}), outflows (\S\ref{sect:outflows}), a broad-line region (BLR; \S\ref{sect:BLR}), an anisotropic dusty absorber (often referred to as the torus; \S\ref{sect:torus}), dust in the polar region of the accreting system (\S\ref{sect:polar}), and a narrow line region (NLR; \S\ref{sect:NLR}). A schematic view of these components, together with their distance from the SMBH and spatial extension, is shown in Figure\,\ref{fig:UM_RR} \citep{Ramos-Almeida:2017ar}.

\subsection{Supermassive Black Hole}\label{sect:SMBH}
\begin{figure*}
\centering
 %% 1st image
 %% 2nd image
\includegraphics[angle=90,width=0.78\textwidth]{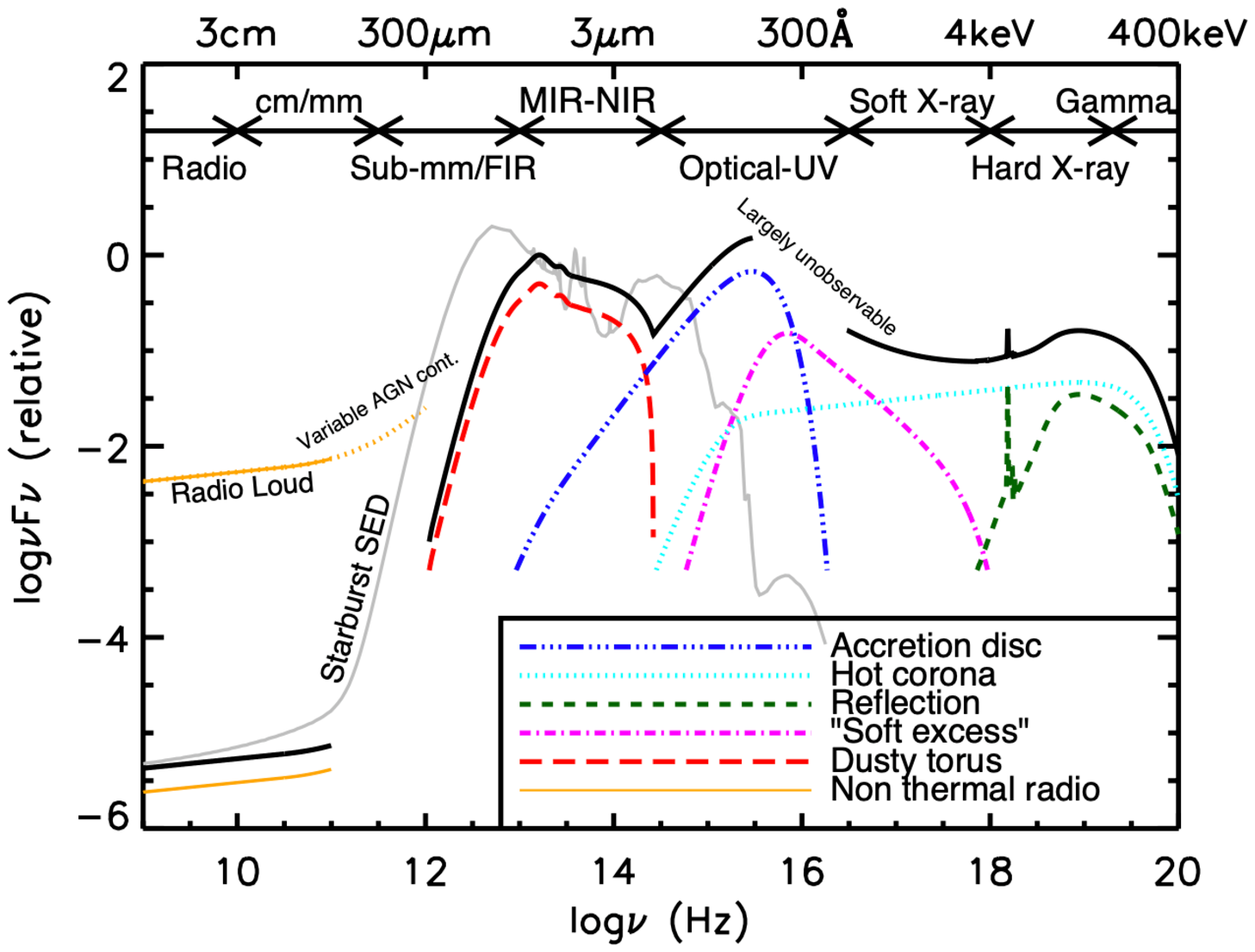}
% %% caption
  \caption{A schematic illustration of the spectral energy distribution (SED) of an unobscured AGN (black curve), displaying its primary physical components (represented by colored curves) and contrasted with the SED of a star-forming galaxy (light gray curve). Figure from \cite{Harrison:2014tv}. }
\label{fig:SEDAGN}
\end{figure*}

At the heart of every AGN lies a supermassive black hole, with masses ($M_{\rm BH}$) typically larger than $10^6 M_\odot$, although lower-mass accreting black holes have been found at the center of nearby dwarf galaxies (e.g. \citealp{Filippenko:2003bn}).  These massive black holes are the {\it engines} of AGNs, powering them through gravitational accretion, and their shadow was recently captured in images obtained by the Event Horizon Telescope (e.g., \citealp{Event-Horizon-Telescope-Collaboration:2019io}). The intense gravitational field of the black hole drives the infall of matter, forming an accretion disk that generates large amount of energy, allowing the system to overshine the host galaxy across the whole electromagnetic spectrum (see Figure\,\ref{fig:SEDAGN}; e.g., \citealp{Elvis:1994wl,Hickox:2018jh}). Studies of large and well-defined samples of AGN, together with accurate measurements of $M_{\rm BH}$ and luminosities, can provide luminosity, Eddington ratio and black hole mass functions (e.g., \citealp{Vestergaard:2009hg,Kelly:2013uy,Ueda:2003qf,Ueda:2014ix,Ananna:2022fg}), which can yield fundamental insights into the demographics, growth history, and evolutionary processes of SMBHs. The duration of the accretion phase in SMBHs is still not well constrained, with studies reporting values between $\sim 10^{5}$ and $10^{8}$\,years (e.g., \citealp{Shen:2007ab,Shankar:2009xo,Schawinski:2015pe}).  It is also still unclear what triggers the accretion process, and what is the role of the host galaxy, with secular processes, such as bars, nuclear star formation, and mergers being among of the main suspects (e.g., \citealp{Sani:2010gi,Davies:2007qo,Treister:2012ys}).

\subsection{X-ray corona}\label{sect:corona}
X-ray emission is ubiquitous in AGN, and is produced in a compact source located within a few gravitational radii from the accretion disk (e.g., \citealp{Fabian:2009dl,Zoghbi:2012ih}; see \citealp{Laha:2025ly} and \citealp{Kara:2025oe} for recent reviews). This corona is thought to consist of very hot, ionized gas that emits X-rays primarily through the process of Compton scattering, where lower-energy (optical/UV) photons from the accretion flow are upscattered to higher energies by the energetic electrons (e.g., \citealp{Haardt:1991qr}). The corona is believed to be powered by the intense magnetic field around the accreting SMBH (e.g., \citealp{Merloni:2001zm,Cao:2009nl}). The Comptonization process results in a power-law spectrum, consistent with what is commonly observed in nearby AGN \citep{Mushotzky:1993de}, with a characteristic cutoff at high energies determined by the temperature of the coronal electrons. The typical photon index of this primary X-ray emission is $\Gamma \sim 1.8$ (e.g., \citealp{Mushotzky:1993de,Ricci:2017pm}), while the median temperature of the Comptonizing plasma, obtained from broad-band X-ray studies, is $kT_{\rm e} \sim 100$\,keV \citep{Ricci:2018du}.
The relationship between the X-ray corona and the accretion disk is complex, with variations in accretion rates influencing the coronal properties, leading to changes in the observed X-ray spectrum (e.g., \citealp{Shemmer:2008tj,Trakhtenbrot:2017ng}). The X-ray corona typically emits $\sim 5\%$ of the bolometric output of AGN (e.g., \citealp{Vasudevan:2007fk,Gupta:2024xz}), and this fraction typically decreases as the AGN luminosity and Eddington ratio increase (e.g., \citealp{Vasudevan:2009zh,Lusso:2012jc,Duras:2020mk,Gupta:2024xz,Gupta:2025vr}). Located near the SMBH, the X-ray corona can provide very valuable insights into the structure and properties of the circumnuclear material through the analysis of both absorbed (e.g., \citealp{Ueda:2014ix,Ricci:2015fk}) and reprocessed (e.g., \citealp{Matt:1991tu}) X-ray radiation. In particular, studying reprocessed X-ray radiation can reveal important details about the structure and physical characteristics of the material around SMBHs, including the accretion flow (e.g., \citealp{Fabian:2000cj}).

\subsection{Accretion Disk}\label{sect:disk}

When matter falls toward a compact object, its gravitational potential energy decreases, converting into kinetic energy. However, conservation of angular momentum prevents the material from directly plunging into the compact object, leading instead to the formation of an accretion disk. In active galactic nuclei (AGN), gas is funnelled toward the supermassive black hole (SMBH) via an accretion flow, whose properties depend strongly on the Eddington ratio (\(\lambda_{\rm Edd} \propto L_{\rm bol}/M_{\rm BH}\), where \(L_{\rm bol}\) is the bolometric luminosity). When SMBHs accrete at moderate to high Eddington ratios (\(10^{-2} \lesssim \lambda_{\rm Edd} \lesssim 1\)), the flow is radiatively efficient and could take the form of a Shakura-Sunyaev optically thick, geometrically thin accretion disk. This model, introduced by \citet{Shakura:1973sm}, describes such disks as geometrically thin and optically thick, with matter orbiting the SMBH in nearly Keplerian trajectories. Viscous forces, likely driven by turbulence and magnetic instabilities (e.g., \citealp{Balbus:1998jy,Kato:2008nw}), transport angular momentum outward, allowing material to spiral inward. As the gas moves closer to the SMBH, gravitational energy is converted into heat, which is radiated as blackbody emission at different temperatures depending on the radius. The inner regions, being hotter, emit higher-energy radiation, while the cooler outer regions emit at lower energies. This creates a multi-temperature spectrum with most of the emission in the optical and ultraviolet (UV) bands. Most luminous AGN in the local universe are thought to accrete through such thin disks. These disks are critical to the AGN energy budget, often accounting for more than $95\%$ of the bolometric luminosity in non-jetted AGN (e.g., \citealp{Vasudevan:2007fk,Gupta:2024xz}). Despite their significant contribution to optical and UV radiation, the large mass of SMBHs means thin disks alone cannot produce the strong X-ray emission observed in AGN, requiring the presence of an additional component (i.e., the X-ray corona; \S\ref{sect:corona}). It should be noted that, while the geometrically thin, optically thick accretion disk model is often used as the standard framework for interpreting the optical and ultraviolet emission of AGN, it cannot explain several observational properties of AGN. These include, for example, the amplitude and timescales of optical/UV variability (e.g., \citealp{Alloin:1985aq}), the larger-than-expected emitting sizes \citep[e.g.,][]{Cackett:2018qe}, and the difficulty of reproducing some observed properties of the optical/UV continuum \citep[e.g.,][]{Antonucci:2023ux,Cai:2023cl}. These discrepancies suggest that the structure of the optical/UV emitting region may be more complex than described by the simplest accretion disk models.

Two additional accretion scenarios are commonly invoked to explain accretion at very low and very high Eddington ratios. At $\lambda_{\rm Edd} \lesssim 10^{-2}$, accretion can be described by an optically-thin Advection Dominated Accretion Flow (ADAF;  \citealp{Narayan:1994cd}), which radiates weakly and lacks significant optical or UV emission. In this scenario, the gas density is too low to cool efficiently within the accretion timescale. Consequently, viscous energy is retained as thermal energy and advected into the central object. This model is also referred to as the two-temperature ADAF, since ions are expected to be significantly hotter than electrons. In SMBHs accreting very rapidly ($\lambda_{\rm Edd}\gtrsim 1$) the accretion flow is also expected to be different from the typical Shakura-Sunyaev optically-thick geometrically-thin accretion disk. Early theoretical works (e.g., \citealp{Abramowicz:1988ng}) showed that  at high Eddington ratios SMBHs are believed to accrete through slim disks. In these objects, the large optical depth of the inflowing gas traps a significant fraction of the radiation, advecting it into the central black hole, and decreasing the overall radiative efficiency of the accretion process. More recently, three dimensional radiation magneto-hydrodynamic simulations (e.g., \citealp{Jiang:2014ay}) have shown that powerful outflows are expected at these accretion rates.  Theoretical works have also shown that at very high-Eddington ratios AGN might display very different spectral energy distributions from lower $\lambda_{\rm Edd}$ AGN, and emit very strongly in the extreme UV and soft X-rays (e.g., \citealp{Kubota:2019sh}).

\subsection{Relativistic Jets}\label{sect:jets}

In some AGNs, relativistic jets of charged particles are ejected, likely along the black hole's rotational axis. Jets can extend far beyond the host galaxy, and in some cases they have been observed Mpc away from their host galaxies (e.g., \citealp{Clarke:2017be}). Jets are associated with radio-loud AGNs and are a prominent feature of two classes of AGN called blazars and radio galaxies (see \S\ref{sect:AGNclassification}). The orientation of these jets with respect to the observer's line of sight influences the observed brightness and variability due to Doppler boosting and relativistic beaming (e.g., \citealp{Urry:1995ga}). Jets can produce large radio lobes that extend beyond their host galaxies, and which are particularly clear in some radio galaxies (e.g., \citealp{Wilson:1995vs}).  Jets mainly emit non-thermal radiation via synchrotron and inverse-Compton scattering processes (e.g., \citealp{Blandford:2019fc}), and are responsible for a significant portion of the observed radio and gamma-ray emission from radio-loud AGN (e.g., \citealp{Celotti:2008hm}). Over the past few decades, advancements in radio interferometry have enabled the detection of jets in numerous sources and even allowed detailed resolution of their inner regions (e.g, \citealp{Lister:2009wv}). We still lack a clear understanding of how jets are produced, accellerated and collimated. Some of the most commonly-adopted models are based on magnetohydrodynamics (e.g., \citealp{Blandford:2001dm}), and propose that strong electromagnetic fields extract the rotational energy of a spinning black hole and drive an outflow through their interaction with differential rotation. Faint jets, producing non-thermal emission, can be also found in radio-quiet AGN, although they are typically uncollimated and sub-relativistic (e.g., \citealp{Panessa:2019nm}). Jets are thought to play a vital role in galaxy evolution by injecting substantial energy into their surrounding environment, a process referred to as radio-mode feedback (e.g., \citealp{Fabian:2012mk}).

\begin{figure*}
\centering
 %% 1st image
 %% 2nd image
\includegraphics[angle=90,width=0.78\textwidth]{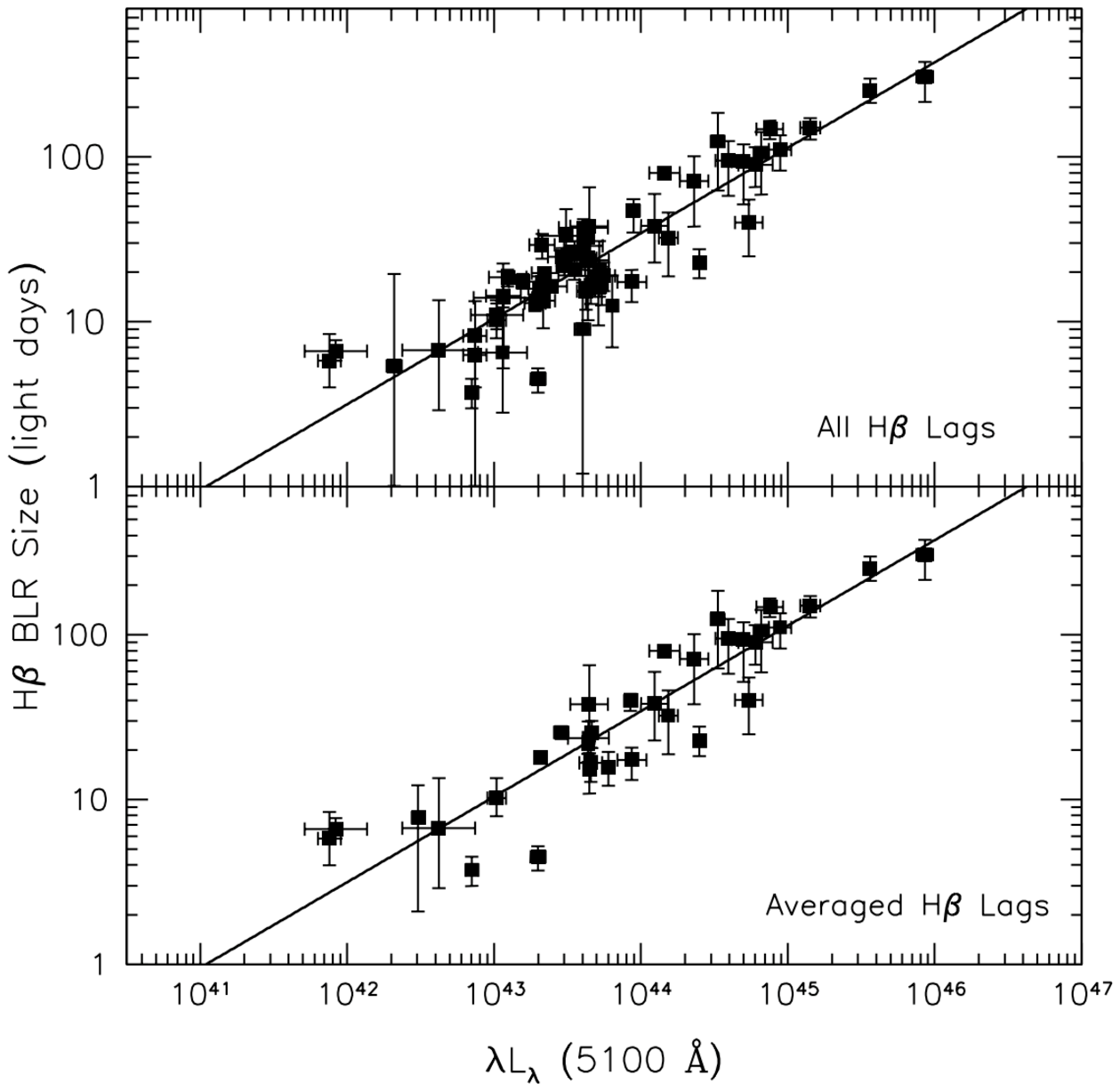}
% %% caption
  \caption{Relationship between the H$\beta$ BLR radius ($R_{\rm BLR}$) and the AGN luminosity ($L$). In the top panel, each measurement is represented as an individual data point, while the bottom panel displays the weighted mean for multiple measurements of the same object. The solid lines represent the best-fit model to the relationship. Figure from \cite{Bentz:2009tc}. }
\label{fig:BLR}
\end{figure*}

\subsection{Outflows}\label{sect:outflows}

A significant fraction of AGN have been found to show outflows, which can be categorized into different phases based on their physical properties and the methods used to detect them.  {\it Warm ionized outflows} are typically detected through optical and near-infrared emission lines such as [O III]$\lambda5007$, and correspond to gas at temperatures $T\sim 10^4$\,K. These outflows are observed on scales ranging from tens of pc to several kpc, with velocities spanning from several hundred to a few thousand $\rm km s^{-1}$ (e.g., \citealp{Harrison:2018zw}). {\it Cold molecular outflows} are typically detected through molecular emission lines, such as for example CO, at millimeter and submillimeter wavelengths, and probe the cold gas phase at temperatures of $\sim 10$\,K. These outflows are generally observed on kiloparsec scales and exhibit velocities of several hundred $\rm km\,s^{-1}$ (e.g., \citealp{Feruglio:2010ct,Veilleux:2013yf,Cicone:2014eb}).  Mildly ionized {\it warm absorbers} with velocities $\sim 500-1000\rm\,km\,s^{-1}$ have been detected through X-ray observations in a large fraction of AGN not showing strong cold absorption features (e.g., \citealp{Piconcelli:2005tg,Blustin:2005fw}). {\it Ultrafast outflows} are observed in X-ray spectra through blue-shifted absorption lines, particularly above 6\,keV. These highly ionized outflows reach relativistic velocities $\sim 0.2-0.4$\,c (e.g., \citealp{Pounds:2003ex,Reeves:2009gg,Tombesi:2010ln}) and are thought to originate near the SMBH.  {\it Optical and UV Broad Absorption Line} systems (e.g., \citealp{Trump:2006xw}), associated with ionized gas outflows at velocities from $2000$ to $20000\rm km\,s^{-1}$, are observed in $20-40\%$ of high-luminosity AGN (e.g., \citealp{Dai:2008mi}). Radio observations have identified numerous neutral {\it atomic gas outflows} with velocities up to $\sim 1500\rm\,km\,s^{-1}$ (e.g., \citealp{Morganti:2005uq}). AGN outflows are thought to be able to significantly quench start formation in their host galaxies, and as such they are believed to be one of the main feedback mechanisms (e.g., \citealp{Fabian:2012mk,King:2015bi}) that influence galaxy evolution.

AGN outflows are believed to be launched through a variety of mechanisms (e.g., \citealp{Proga:2007mt}), such as: i) Thermal driving (e.g., \citealp{Proga:2002oj}). The heating of the accretion disk's upper atmosphere, particularly through X-ray irradiation, can elevate the gas to a temperature of $\sim 10^7 \, \mathrm{K}$. If the thermal velocity exceeds the local escape velocity, this process can drive a thermal wind, primarily effective at larger radii where the escape velocity is lower. ii) Radiation Pressure (e.g., \citealp{Shlosman:1985jh,Ishibashi:2018ya}). In moderately ionized gas, radiation pressure, particularly on spectral lines, can impart significant momentum to the gas. This mechanism is effective when UV line opacity is substantial and can accelerate gas to high velocities, particularly in luminous AGN with strong UV radiation fields. iii) Magnetic Driving (e.g., \citealp{Fukumura:2010tm}): Magnetic fields, integral to the structure of accretion disks, can launch winds through magnetocentrifugal or magnetic pressure-driven processes. Magnetic driving could explain the extreme velocities found in ultra-fast outflows (e.g., \citealp{Kraemer:2018qc}). These three mechanisms are not mutually exclusive and may operate simultaneously in different regions of the accretion disk or under varying physical conditions.

\begin{figure*}
\centering
\includegraphics[angle=90,width=\textwidth]{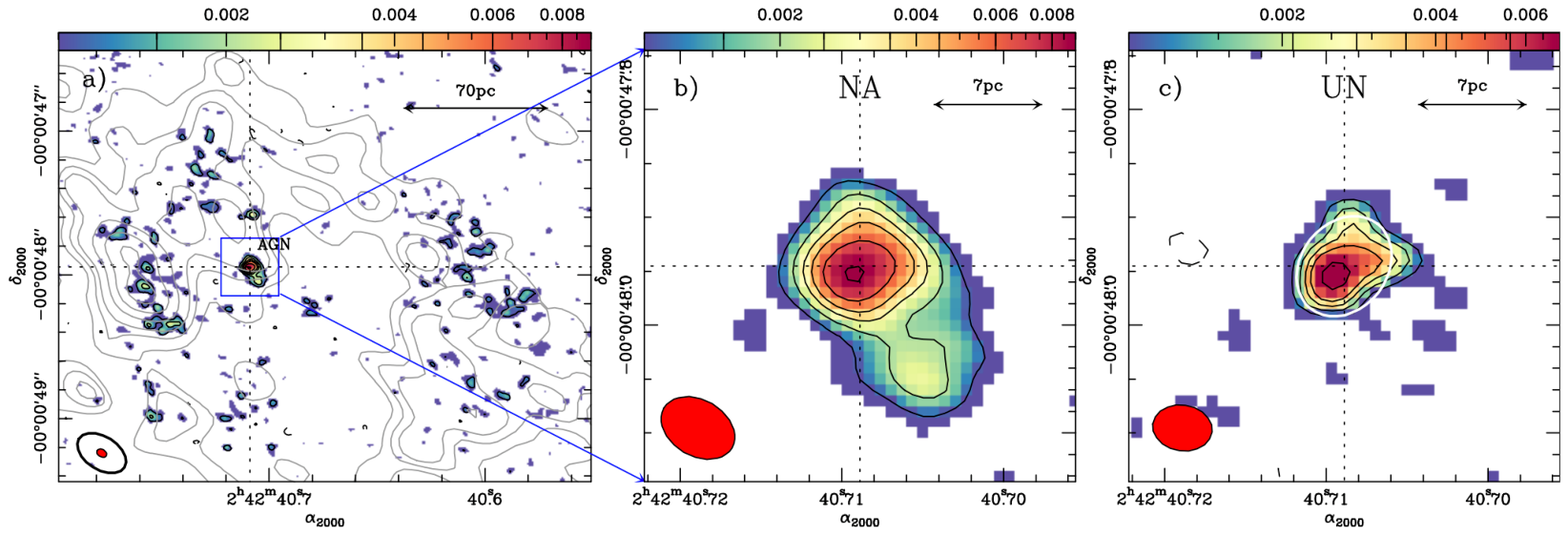}
\vspace{-5cm}
\caption{ALMA Observations of the nuclear region of NGC 1068: the left panel shows the dust continuum emission at 694 GHz, while the central panel show zoomed in views. The right panel is the same as the central one, but using a uniform-weighted set of data. Figure from \cite{Garcia-Burillo:2016uc}.}
\label{fig:torGB16}
\end{figure*}

\subsection{Broad-Line Region}\label{sect:BLR}

Strong broad emission lines are a defining feature of AGN optical/UV spectra. These lines are Doppler-broadened and originate in a photo-ionized region near the black hole, commonly referred to as the broad-line region (BLR). The BLR consists of high-density gas clouds in Keplerian motion, primarily under the gravitational influence of the SMBH. The BLR is thought to be dust-free, as it resides within the sublimation radius, and it emits broad permitted and semi-forbidden emission lines, with velocity widths ranging from 1,000 to 20,000\,$\mathrm{km\,s^{-1}}$. These lines, including prominent transitions such as H$\alpha$, H$\beta$, Ly$\alpha$, C IV, and Mg II, are produced by photoionization as the gas is illuminated by radiation from the accretion flow. The extremely high gas densities in the BLR, typically exceeding $10^9\,\mathrm{cm^{-3}}$, prevent the formation of forbidden lines such as [O III], which are instead observed in the more distant and rarified gas (e.g., \citealp{Netzer:2015cx}). 

Reverberation mapping, which measures the time delay between AGN continuum variations and the corresponding response of broad lines, has allowed to constrain the emissivity-weighted radius of the BLR (e.g., \citealp{Kaspi:2000bv}), which typically ranges from a few to hundreds light days from the SMBH. These measurements have led to the discovery of a strong positive correlation between the size of the BLR and the AGN luminosity  (Figure\,\ref{fig:BLR}; \citealp{Kaspi:2000bv, Bentz:2009tc}). This radius-luminosity relationship is fundamental for estimating SMBH masses, providing a direct link between the BLR dynamics and the gravitational potential of the central engine. This has allowed to estimate black hole masses from single-epoch spectroscopy, directly connecting the width of the broad lines to the black hole mass (e.g., \citealp{Shen:2008ha,Trakhtenbrot:2012sw}). This approach has been applied to a large number of AGN, mostly using H$\beta$ (e.g., \citealp{Vestergaard:2002sn}), H$\alpha$ (e.g., \citealp{Greene:2005mq}) and MgII (e.g., \citealp{McLure:2002ng}). Recent high-resolution observations with GRAVITY have been able to resolve the BLR, showing that it can be well described by a rotating, thick disk with a radial distribution of clouds peaking in the inner region \citep{GRAVITY-Collaboration:2021yf}, and that its size is  consistent with the radius-luminosity relation \citep{GRAVITY-Collaboration:2020wa}. Observations suggest that the BLR is not a homogeneous structure but comprises zones with varying ionization states and densities, reflecting the diverse physical conditions across its extent. Its geometry and kinematics are influenced by AGN properties such as luminosity and orientation, with evidence of rotational motion, inflows, and outflows in some systems (e.g., \citealp{Pancoast:2014zb}).

\begin{figure*}
\centering
\includegraphics[angle=90,width=0.7\textwidth]{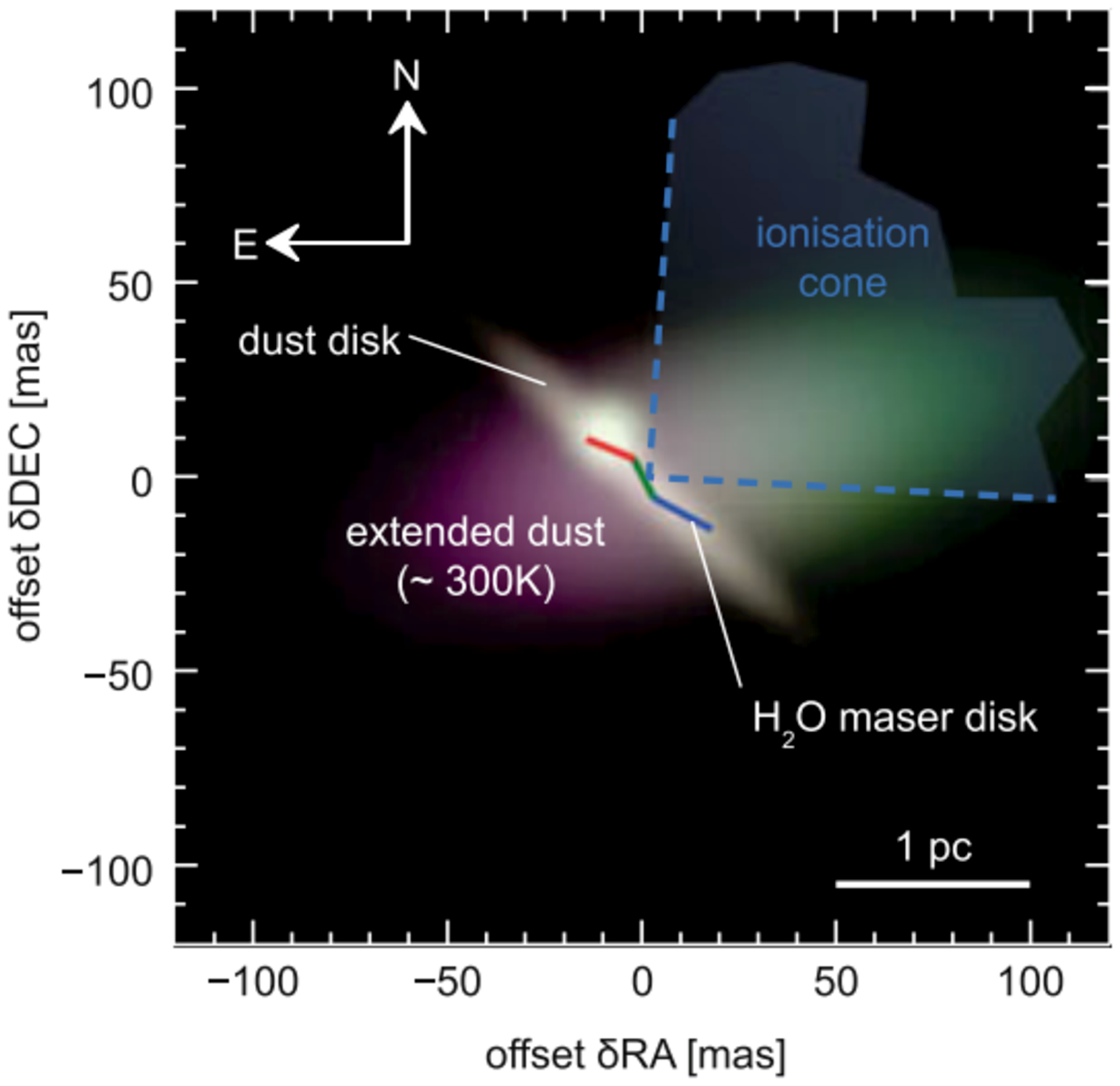}
\includegraphics[angle=90,width=0.7\textwidth]{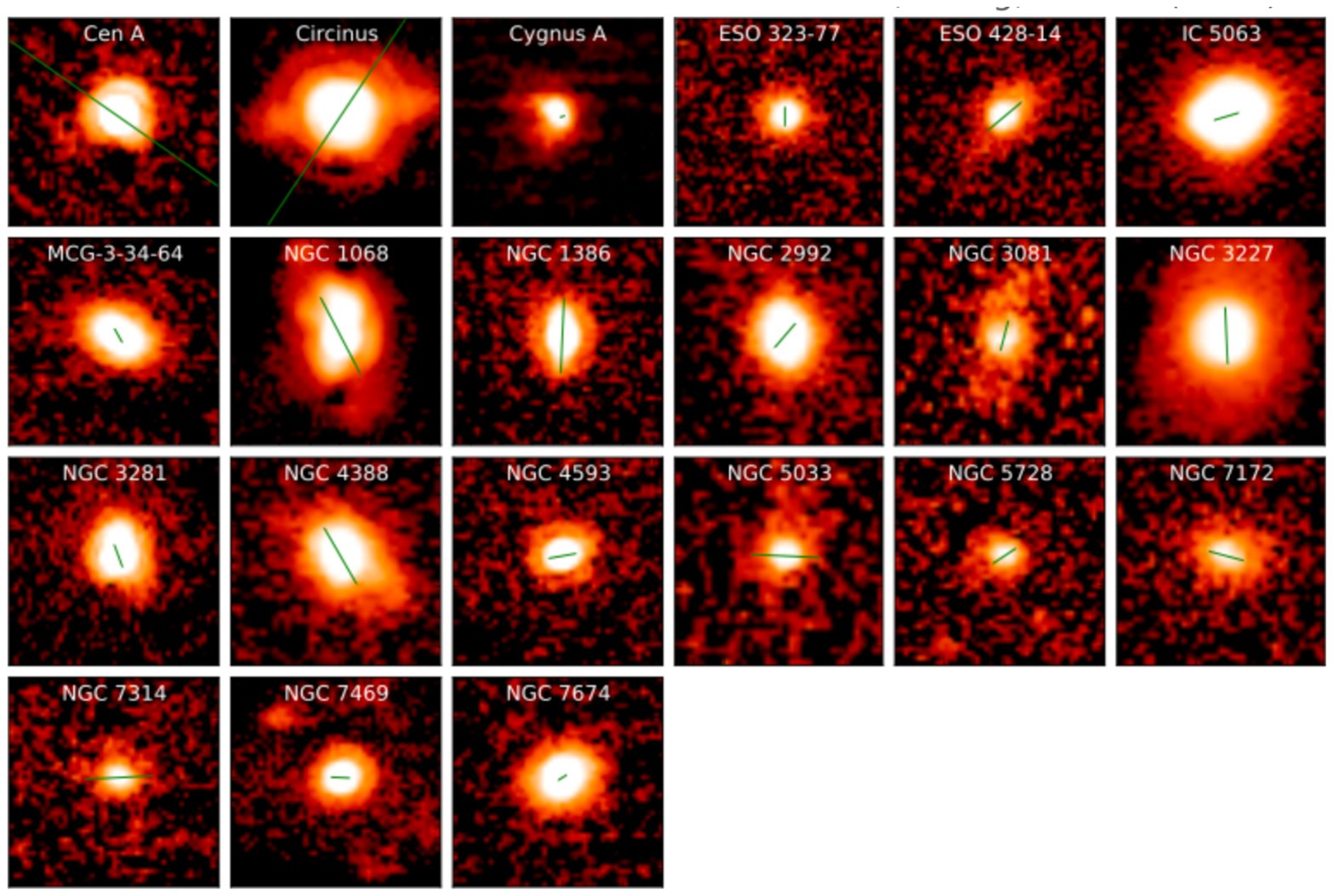}
\caption{{\it Top panel}: mid-infrared emission from the nucleus of the Circinus galaxy obtained from mid-IR interferometric observations and different components. Figure from \cite{Tristram:2014ss} and \cite{Honig:2016ko}. {\it Bottom panel:} 
Mid-infrared images of the nuclear regions of AGN showing extended mid-infrared emission. The green lines indicate the position angle of the system axis, with each line measuring 100 pc in length. Figure from \cite{Asmus:2016uf}.}
\label{fig:polardust}
\end{figure*}

\subsection{Dusty anisotropic obscurer}\label{sect:torus}

One key ingredient of AGN, fundamental for the different AGN classifications, is the dusty anisotropic obscurer, also called the torus (e.g., \citealp{Antonucci:1993vt,Ramos-Almeida:2017ar,Honig:2019lt}). This torus obscures the inner regions (including the BLR and accretion disk) when viewed edge-on. In contrast, a face-on view provides a clear line of sight to the BLR and accretion disk, revealing broad emission lines. The torus plays a pivotal role in the unification model by creating the orientation-dependent obscuration (see \S\ref{sect:AGNclassification}). Traditionally conceptualized as a geometrically thick and optically dense structure surrounding the accretion disk, the torus is composed of dust and gas that absorbs and reprocesses the radiation from the central engine, emitting strongly in the infrared. The infrared radiation produced in this component is tightly correlated with the X-ray and bolometric AGN luminosity (e.g., \citealp{Gandhi:2009dq}). The torus extends from the dust sublimation radius (e.g., \citealp{Kishimoto:2009pc}), and it has been found to have typical sizes of 0.1$-$10 pc from mid-IR imaging (e.g., \citealp{Packham:2005sr}) and interferometry (e.g., \citealp{Jaffe:2004fj, Burtscher:2013xu}), and more recently from sub-millimiter observations (Figure\,\ref{fig:torGB16}; e.g., \citealp{Imanishi:2016ur,Garcia-Burillo:2016uc,Alonso-Herrero:2018ek,Impellizzeri:2019nj,Goesaert:2025xx}). This structure also collimates the AGN radiation, hence producing the bi-conical shapes of their narrow-line region (\S\ref{sect:NLR}) known as ionization cones (e.g., \citealp{Malkan:1998fn}).  The structure of this component is expected to be clumpy rather than uniform (e.g., \citealp{Nenkova:2008op}), as supported by observations and theoretical models, allowing for partial transmission of light even in AGN observed edge on (e.g., \citealp{Elitzur:2006wq}). Clumpy torus models have been shown to be able to well reproduce the observed infrared spectral energy distributions of AGN (e.g., \citealp{Ramos-Almeida:2009jg,Alonso-Herrero:2011tz}).

Mid-infrared interferometry has been instrumental in resolving the torus structure in nearby AGN. Interferometric observations have revealed a two-component model, with a compact, disk-like structure in the equatorial plane surrounded by a more extended polar dust emission region (e.g., \citealp{Honig:2013wf,Burtscher:2013xu,Tristram:2014ss}, see \S\ref{sect:polar}). The equatorial disk aligns with the classical torus, while the polar component suggests outflows carrying dusty material away from the nucleus. Theoretical models suggest that the torus is a dynamic structure, maintained by the interplay of inflowing and outflowing material \citep{Wada:2012rp,Wada:2016ef}. Radiation pressure from the accretion disk and winds, combined with turbulence and magnetic forces, might contribute to the vertical support of the torus, allowing it to sustain its geometrically thick shape. The torus could also be central to AGN feedback mechanisms (e.g., \citealp{Ricci:2017rn}). As the reservoir of gas and dust of the accretion process, this component regulates the material available for accretion onto the black hole while simultaneously participating in outflows that enrich the surrounding interstellar medium.

\subsection{Polar dust}\label{sect:polar}

Polar dust has emerged as a significant component in the study of AGN, particularly in understanding their infrared  emissions (see e.g., \citealp{Ramos-Almeida:2017ar} for a recent review).  Over the past few years, interferometric studies (e.g., \citealp{Jaffe:2004fj,Honig:2013wf,Burtscher:2013xu,Tristram:2014ss,Gamez-Rosas:2022gp}; left panel of Figure\,\ref{fig:polardust}) and high-spatial resolution observations carried out by 8--10\,m telescopes (e.g., \citealp{Asmus:2016uf,Garcia-Bernete:2016pf}; right panel of Figure\,\ref{fig:polardust}) have shown that a significant fraction of the mid-IR emission of AGN is elongated in the polar direction, and models including a disk and a wind can reproduce the near- to mid-IR properties of local AGN \citep{Honig:2017wm}. The elongated mid-IR emission, which is thought to originate in a hollow cone (e.g., \citealp{Stalevski:2017qv,Stalevski:2019gl}), could be related to dusty outflows associated to radiation pressure (e.g., \citealp{Honig:2012hq,Leftley:2019ib,Venanzi:2020gx}). Such outflows could also be associated to larger-scale winds observed in AGN (see \S\ref{sect:outflows}).  Hydrodynamic simulations incorporating radiation pressure and AGN feedback have demonstrated that such outflows can form transient dust structures aligned along the polar axis \citep{Wada:2012rp,Wada:2016ef}. Observationally, the polar MIR emission has been found to be brighter than the compact torus in many AGN, implying that this dust contributes significantly to the overall infrared luminosity \citep{Honig:2013wf}. This emission is spatially correlated with ionization cones and the narrow-line region (\S\ref{sect:NLR}), as shown in studies of nearby AGN like NGC 1068, where the polar emission extends up to tens of parsecs \citep{Lopez-Gonzaga:2014rm}.

The presence of polar dust could explain the weak differences in MIR emission between different AGN types, which are observed at different inclination angle with respect to the anisotropic obscurer (see \S\ref{sect:AGNclassification}). While clumpy torus models have been proposed to account for this, polar dust emission provides an alternative explanation that aligns with observations of a tight correlation between MIR and hard X-ray luminosities in both obscured and unobscured AGN \citep{Asmus:2015ly}. The isotropic nature of the polar emission, which is less affected by viewing angle, supports the hypothesis that this component is a prominent contributor to the MIR output.

\subsection{Narrow-line region}\label{sect:NLR}

The narrow-line region (NLR) is typically the most extended component of AGN and has been routinely resolved by optical observations (e.g., \citealp{Fischer:2013nk}). In contrast to the BLR, the NLR has a lower electron density, allowing many forbidden transitions to occur without being collisionally suppressed. Some of the most prominent narrow lines are Ly$\alpha\,\lambda1216$, C\,IV\,$\lambda$1549, C\,III]\,$\lambda$1909, [O\,III]\,$\lambda$4959, [O\,III]\,$\lambda$5007, [N\,II]\,$\lambda$6584 and [SII]\,$\lambda$6717 (e.g., \citealp{Heckman:1981bc,Boroson:1992dq}). Narrow emission lines typically have full width at half maximum  values between $200$ and $900 \, \mathrm{km \, s^{-1}}$, with most lines falling within the range of $350$--$500 \, \mathrm{km \, s^{-1}}$. The NLR is normally found to be axisymmetric, and the ratios of some of the lines produced in the NLR can be used as a tracer of AGN activity (e.g., \citealp{Kewley:2006uq}). Similar to the BLR (Figure\,\ref{fig:BLR}), the distance of the NLR from the central engine scales with AGN luminosity, following the relationship $R_{\rm NLR} \propto L^{0.5}$ (e.g., \citealp{Peterson:1997wa}). In nearby and moderately luminous AGN, the NLR is typically located at distances of $\sim 100$--$300 \, \mathrm{pc}$, whereas in more luminous AGN, the NLR may extend to scales of several kiloparsecs. In some cases, the NLR has been found to be very large, while the AGN is either undetected or considerably fainter than expected based on the NLR emission (e.g., \citealp{Lintott:2009vn,Keel:2012vb}). In such objects, the NLR is thought to represent an echo of a previous rapid growth episode.

\section{AGN classification and angle-dependent Unification}\label{sect:AGNclassification}

Over the past 50 years a large number of AGN have been discovered, showing different observational characteristics, which has led them to be divided into different {\it phenomenological} categories. One of the main divisions is related to the presence or not of strong jets: sources without powerful jets, which make up $\sim 90\%$ of the AGN population, are usually called non-jetted or radio-quiet, while those displaying strong jet emission are dubbed radio-loud or jetted. Historically, non-jetted AGN have been often referred to as Seyfert galaxies or quasars, depending on whether the luminosity of the accreting source is low (i.e., $< 10^{44}\rm\,erg\,s^{-1}$) or high (i.e., $> 10^{44}\rm\,erg\,s^{-1}$), respectively. 

\subsection{Non-jetted AGN unification}\label{sect:nonjettedUM}
\begin{figure*}
\centering
\includegraphics[width=0.48\textwidth]{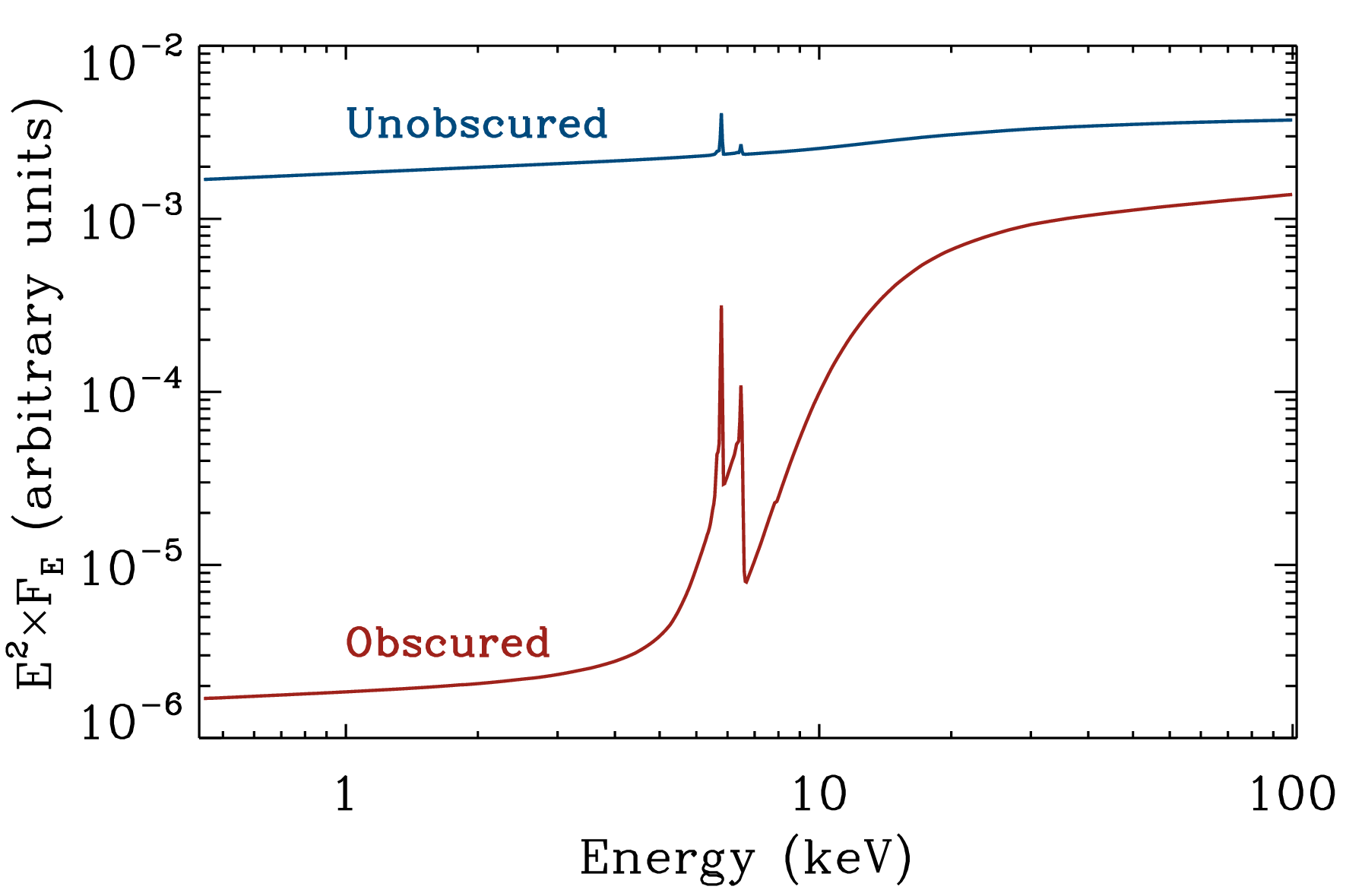}\hfill
\includegraphics[width=0.48\textwidth]{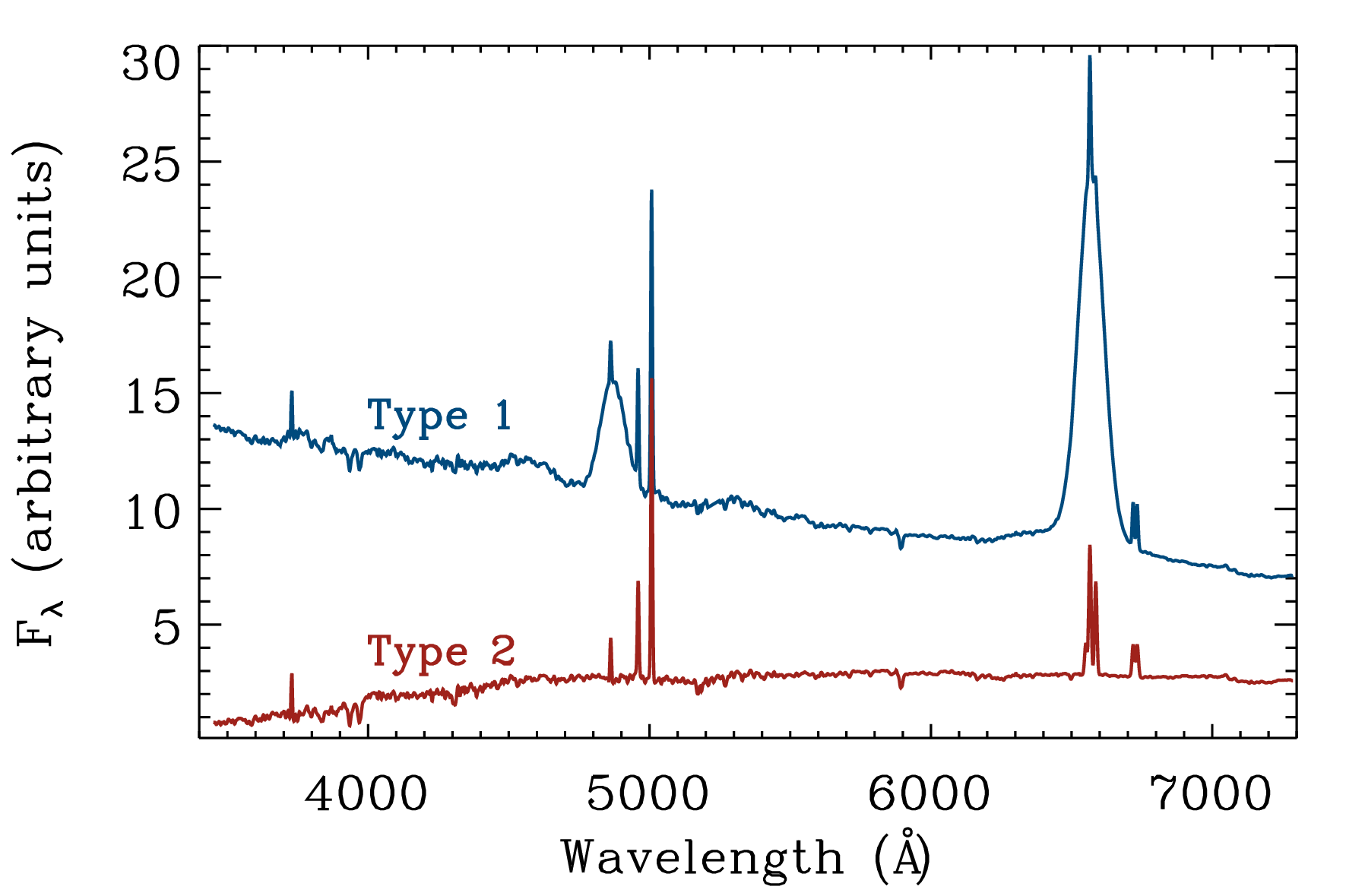}
\caption{X-ray and optical spectra of different classes of non-jetted AGN. {\it Left panel}: X-ray unobscured (blue line) and X-ray obscured (red line) AGN. During a changing-obscuration event (see \S\ref{sect:COAGN}) AGN are found to transition from one class to the other, sometimes reaching Compton-thick column densities. b) Optical Type\,1 (blue line) and Type\,2 AGN (red line). During changing-state events (see \S\ref{sect:CSAGN}) AGN can transition from one class to the other. Figure from \cite{Ricci:2023fw}.}
\label{fig:schematicCLAGN_spec}
\end{figure*}

Non-jetted AGN are usually classified based on the presence/absence of broad optical/UV lines from the BLR, and on the presence/absence of strong line-of-sight absorption associated to neutral material in the X-ray band. X-ray absorption is described by the line-of-sight column density ($N_{\rm H}$) and according to its value AGN can be broadly divided into two categories: unobscured ($N_{\rm H}<10^{22}\rm\,cm^{-2}$) and obscured ($N_{\rm H}\geq10^{22}\rm\,cm^{-2}$). Obscured sources can be further categorized into Compton-thin ($10^{22}\leq N_{\rm H}<10^{24}\rm\,cm^{-2}$) and Compton-thick ($N_{\rm H}\geq10^{24}\rm\,cm^{-2}$). Based on these properties the two main types of non-jetted AGN are the following:
\smallskip

\noindent{\bf Type\,1/unobscured AGNs}. Type\,1 AGNs show strong broad emission lines in the optical/UV, with widths ranging from $1,000$ to $20,000\rm\,km\,s^{-1}$, originating from permitted and semi-forbidden transitions. The spectra of Type\,1 AGNs often feature high-ionization narrow emission lines, many of which correspond to forbidden transitions. Type\,1 AGN can be further divided into Type 1.2, 1.5, 1.8, and 1.9, based on the relative strengths of the broad and narrow components of Balmer lines. For example, Type 1.9 AGNs exhibit a dominant narrow-line component. Type-1 AGNs are typically classified as unobscured in the X-rays (e.g., \citealp{Ricci:2017pm,Koss:2017wu}), and one can observe the primary X-ray radiation produced by the corona. The X-ray and optical spectra of these objects are shown with blue lines in Figure\,\ref{fig:schematicCLAGN_spec}.
\smallskip

\noindent{\bf Type\,2/obscured AGNs}. Type\,2 AGNs are distinguished by narrow emission lines with widths between $300$ and $1,000$~km~s$^{-1}$, which are broader than those typically observed in star-forming galaxies of similar type. These emission lines, such as [O III], [N II], [O II], [O IV], [Ne V], and hydrogen Balmer and Lyman series lines, are indicative of photoionization driven by an AGN. In the X-rays they typically show strong absorption, and are classified as obscured (e.g., \citealp{Awaki:1991xq,Ricci:2017pm,Torres-Alba:2021kl}). The spectra of these objects are shown in red lines in Figure\,\ref{fig:schematicCLAGN_spec}. Large X-ray studies have shown that in the local Universe $\sim 70\%$ of the AGN are X-ray obscured, and $\sim 30\%$ are Compton-thick \citep{Ricci:2015fk,Boorman:2024pi}. The column density distributions of nearby AGN from recent studies are illustrated in Figure\,\ref{fig:NHdistribution}, which shows how most AGN are found to belong to this class. A significant fraction of Compton-thick AGN is also required by synthesis models of the Cosmic X-ray background (e.g., \citealp{Gilli:2007yp,Treister:2009ma,Ueda:2014ix,Ananna:2019rz,Gerolymatou:2026ff,Gerolymatou:2025km})
\smallskip

\begin{figure*}
\centering
 %% 1st image
 %% 2nd image
\includegraphics[width=0.78\textwidth, angle=270]{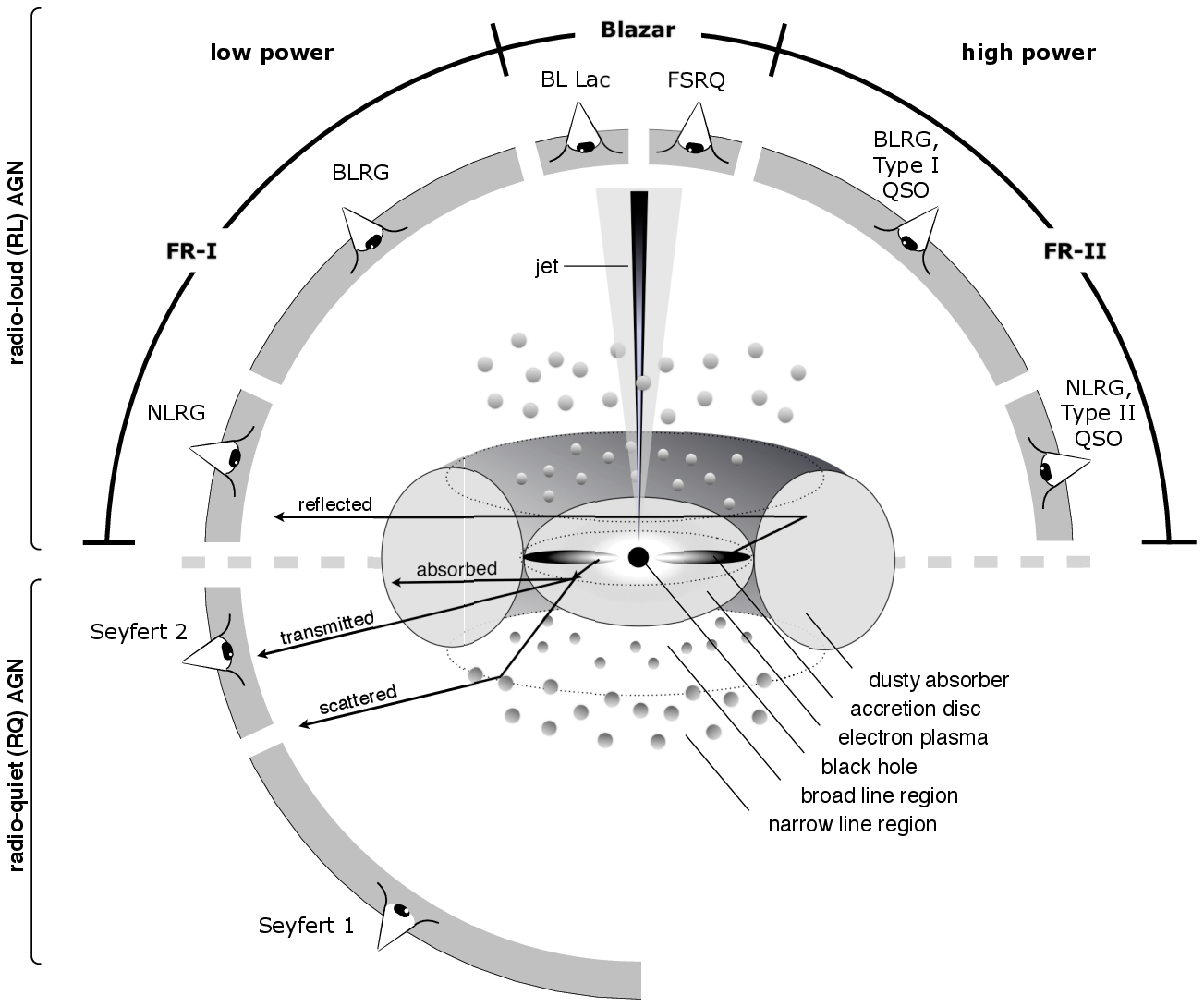}
% %% caption
  \caption{Schematic representation of different classes of jetted (upper part) and non-jetted (lower part) AGN, in which the classification depends on the viewing angle, on the presence of jets, and on the power of the accreting system. Seyfert\,1 and Seyfert\,2 galaxies correspond to Type\,1 and Type\,2 AGN, respectively. Figure from \cite{Beckmann:2012ts,Beckmann:2012fm}.}
\label{fig:UM_RR2}
\end{figure*}

\noindent The discovery of a polarized H$\alpha$ broad-line component in Type\,2 AGNs, with a position angle perpendicular to the radio axis \citep{Antonucci:1984dw} played a key role in shaping the angle-dependent AGN unified model (e.g., \citealp{Antonucci:1993vt}). In this model the main difference between Type\,1 and Type\,2 AGN (or between obscured and unobscured sources) is the inclination angle with respect to the torus at which one observes the accreting SMBH (see lower part of Figure\,\ref{fig:UM_RR2}). In Type\,2/obscured AGNs, the observer is looking at the AGN edge-on, so that the torus blocks direct emission from the BLR and accretion disk, allowing to observe in the optical only the narrow lines produced by more distant photoionized gas in the NLR. The circumnuclear dust also obscures the X-ray source, leaving clear imprints in the X-ray spectrum. In this scheme photons from the BLR scattered by the circumnuclear material produce the broad lines observed in polarized light. In Type\,1/unobscured AGN on the other hand the observer is looking at the accreting source pole-on, so that the BLR, disk and corona can be all observed directly.

\subsection{Jetted AGN unification}
AGN showing powerful jet emission are usually classified into two main categories depending on their multi-wavelength properties:

\smallskip
\noindent{\bf Blazars}. Blazars are characterized by extreme variability, high polarization, and intense radio and gamma-ray emissions. Blazars are further divided into: Flat Spectrum Radio Quasars (FSRQs) and BL Lacertae objects (BL Lacs). FSRQs are distinguished by the strong, broad emission lines in their optical spectra, indicating significant ionization and high luminosities. In contrast, BL Lacs exhibit much weaker or even absent emission lines ($EW<5\AA$), and are generally less luminous than FSRQs (e.g., \citealp{Fossati:1998uo,Giommi:2012jm}).  The two subclasses also exhibit different variability patterns; BL Lacs tend to show a bluer-when-brighter trend, while FSRQs typically exhibit a redder-when-brighter behavior \citep{Gu:2006qc}.

\smallskip
\noindent{\bf Radio Galaxies} Radio galaxies are characterized by their strong radio emissions, which are typically associated with relativistic jets from their central SMBHs. They are primarily classified into two categories based on their radio luminosity and morphology: Fanaroff-Riley type I (FR I) and type II (FR II) radio galaxies. FR I radio galaxies exhibit one-sided jets and are generally less powerful, while FR II radio galaxies display more powerful, two-sided jets and are often associated with higher luminosities \citep{Fanaroff:1974nf}. Similar to non-jetted AGN, radio galaxies may or may not display broad lines and can be classified as broad-line or narrow-line radio galaxies, respectively.

\smallskip

\noindent The unification model of jetted AGN (e.g., \citealp{Urry:1995ga}; see top panels of Figure\,\ref{fig:UM_RR}) posits that blazars are radio galaxies for which the relativistic jet is oriented close to the observer's line of sight, leading to strong relativistic beaming effects that amplify their brightness and variability. Differences between broad and narrow line radio galaxies are expected to be explained in the same way as non-jetted AGN, i.e. with the orientation with respect to the anisotropic dusty obscurer. In this scheme, FR\,I radio galaxies  and BL Lacs are intrinsically the same population of low-power jetted AGN, viewed from different inclination angles with respect to the powerful jet. Similarly, FSRQs are the Doppler boosted counterparts of FR\,II radio galaxies, with both of them being high-power jetted AGN. The FSRQ/FR\,II and BL Lac/FR\,I dichotomy could be attributed to different accretion modes (e.g., \citealp{Prandini:2022ul}), with the former powered by optically thick, geometrically thin disks, and the latter by radiatively inefficient ADAFs (see \S\ref{sect:disk}).

\section{The covering factor of the obscuring material and radiation-regulated unification}\label{sect:RRunification}

One of the key parameters describing the obscuring material around SMBHs is its covering factor, which quantifies the fraction of the SMBH's sky that is obscured (e.g., \citealp{Lawrence:2010ov}). In the simple inclination-angle-dependent unification model (\S\ref{sect:nonjettedUM}), the likelihood of observing an AGN as Type\,2 or obscured depends solely on the inclination angle relative to the dusty anisotropic obscurer. A large covering factor increases the probability of observing an AGN as obscured, while a small covering factor makes it more likely to observe the AGN as unobscured. Thus, the covering factor is a critical element in unification models, and any systematic variation of this quantity with AGN physical parameters could significantly impact unification. This parameter can be estimated using several approaches across different energy regimes:\newline

\begin{itemize}
\item \textit{Surveys:} In X-ray surveys, the covering factor can be inferred from the fraction of obscured sources ($f_{\rm obs}$), i.e. the fraction of AGNs with column densities $\log (N_{\rm H}/\mathrm{cm}^{-2}) \geq 22$. This approach relies on the unified model, where the observed population probes different inclination angles relative to the torus. A complete sample (or a sample with a well-defined selection function) allows the fraction of sources with specific column density ranges to act as a proxy for the mean covering factor of the obscuring material \citep[e.g.,][]{Ueda:2014ix,Merloni:2014iw,Ricci:2015fk}. Similarly, the covering factor can also be derived from optical surveys by comparing the fraction of AGNs classified as Type\,1 and Type\,2 \citep[e.g.,][]{Lawrence:1982bt,Toba:2013jf}. However, this method is not as reliable as the one that considers X-ray surveys, with additional uncertainties arising from the presence of optically dull AGNs \citep[e.g.,][]{Smith:2014pf}, and unobscured AGNs with faint or undetectable broad optical lines \citep[e.g.,][]{Bianchi:2017oj}.\newline
\item \textit{Infrared to Bolometric Luminosity Ratio:} Since a significant portion of IR emission from AGNs originates from dust reprocessing, the ratio of the IR luminosity to the total bolometric luminosity can be assumed to be proportional to the covering factor of the dust \citep[e.g.,][]{Maiolino:2007ii,Treister:2008ff}. However, detailed radiative transfer simulations of dusty tori have revealed that this relationship is more nuanced than initially assumed, and corrections associated to the anisotropic emission of accretion flow and circumnuclear dust are needed \citep{Stalevski:2016kl}. Moreover, the presence of a strong infrared emission component from polar dust might complicate this further.\newline
\item \textit{Mid-IR Spectral Energy Distribution:} Modeling the mid-IR SED of AGNs provides another method to estimate the properties of the obscuring dust, including the covering factor. Various torus models have been developed \citep{Honig:2017wm,Stalevski:2012kq,Stalevski:2016kl}, and applied to numerous AGN datasets \citep[e.g.,][]{Alonso-Herrero:2021yl,Ramos-Almeida:2011eb,Ichikawa:2015qq,Garcia-Bernete:2019mp,Garcia-Bernete:2022pk}.\newline
\item \textit{X-ray Spectral Modeling:} Broad-band X-ray spectra of AGNs also provide insights into the covering factor of the surrounding gas and dust. Recent torus models \citep[e.g.,][]{Paltani:2017zt,Balokovic:2018nb,Tanimoto:2019ts,Buchner:2021kv,Ricci:2023lp,Vander-Meulen:2023uz,Dimopoulos:2025fc,Fujiwara:2026dp} have been applied to broad-band X-ray spectra to infer covering factors of nearby AGN \citep[e.g.,][]{Ogawa:2019jz,Zhao:2020ql,Yamada:2021vz,Andonie:2022qx}.\newline
\end{itemize}

\begin{figure*}
\centering
 %% 1st image
 %% 2nd image
\includegraphics[angle=90,width=0.48\textwidth]{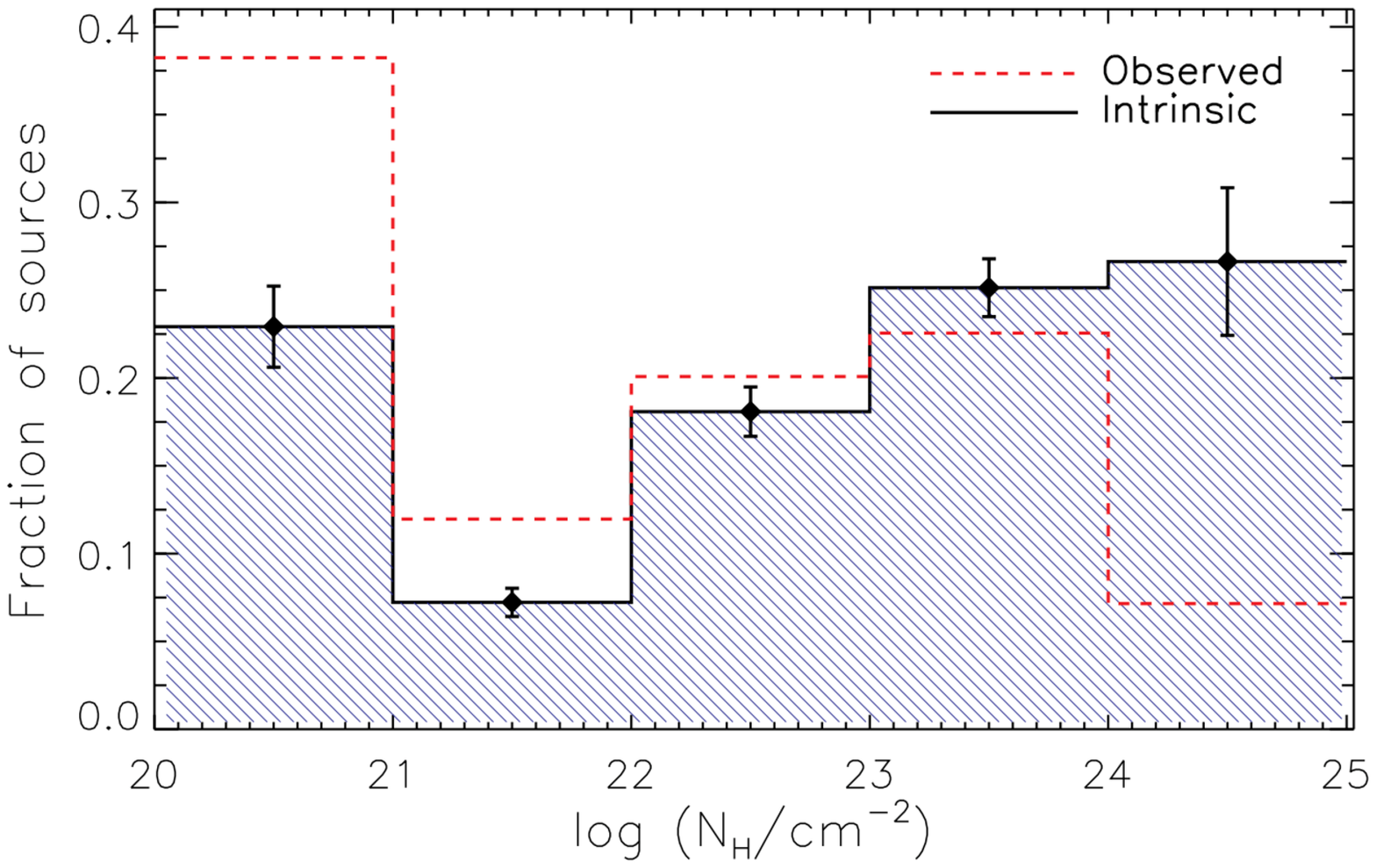}
\includegraphics[angle=90,width=0.48\textwidth]{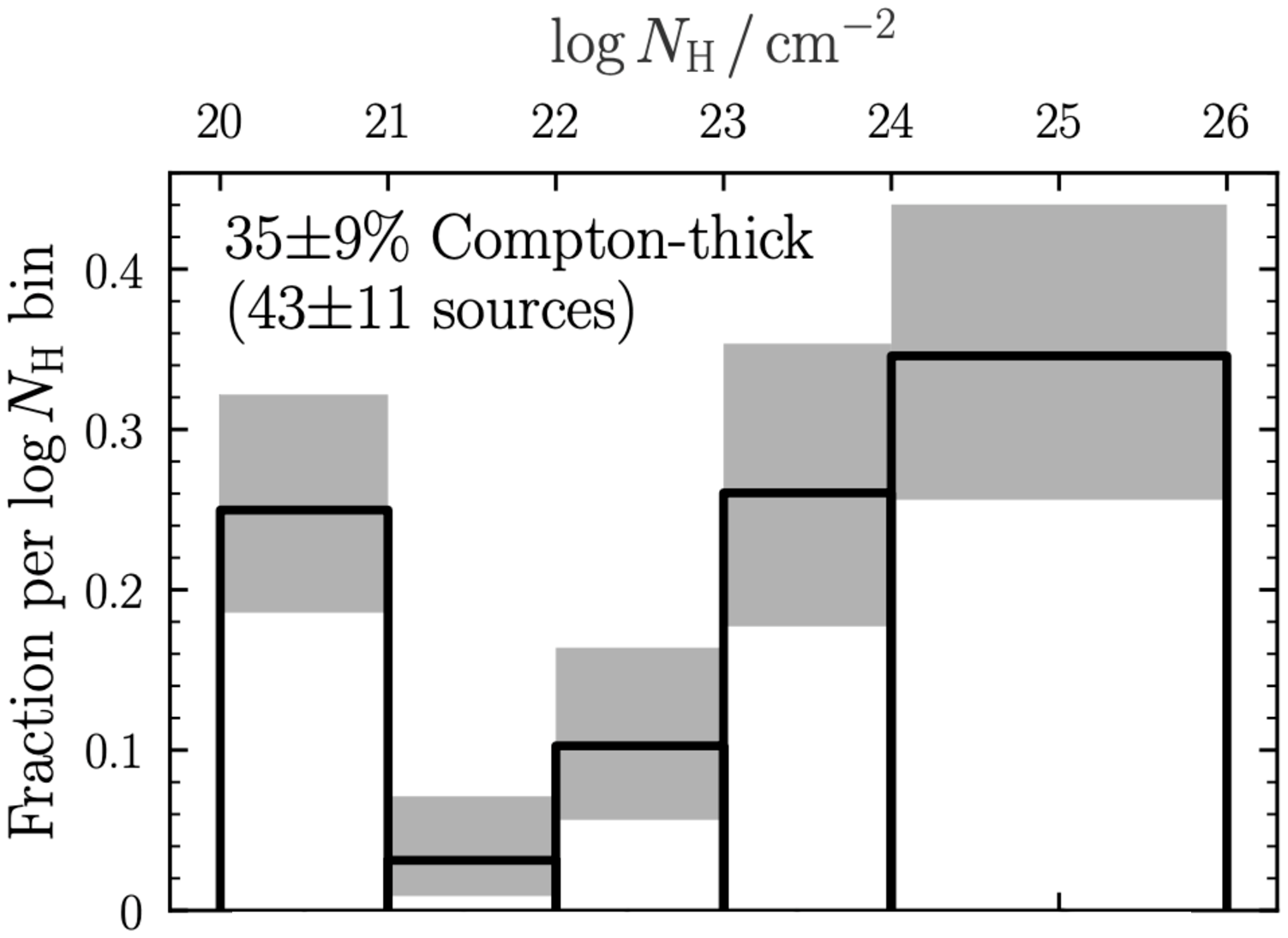}
% %% caption
  \caption{Column density distribution obtained by large X-ray studies of nearby AGN. {\it Left panel}: $N_{\rm H}$ distribution of AGN from the hard X-ray ($>10$\,keV) selected Swift/BAT sample \citep{Ricci:2017pm}. {\it Right panel}: $N_{\rm H}$ distribution for a far-infrared selected sample \citep{Boorman:2024pi}.}
\label{fig:NHdistribution}
\end{figure*}

\begin{figure}[t!]
 \centering
\includegraphics[width=0.48\textwidth]{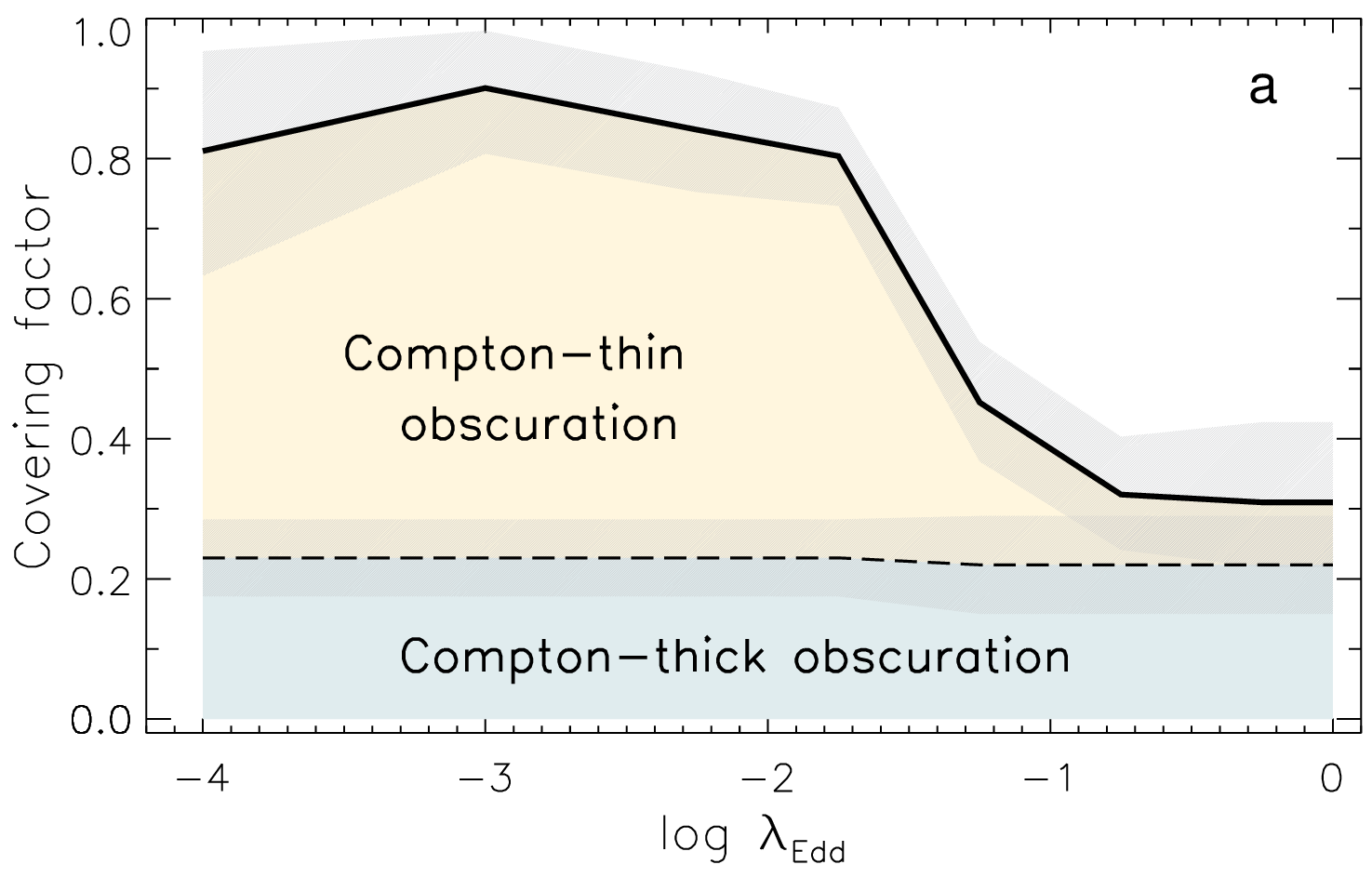}
\includegraphics[width=0.48\textwidth]{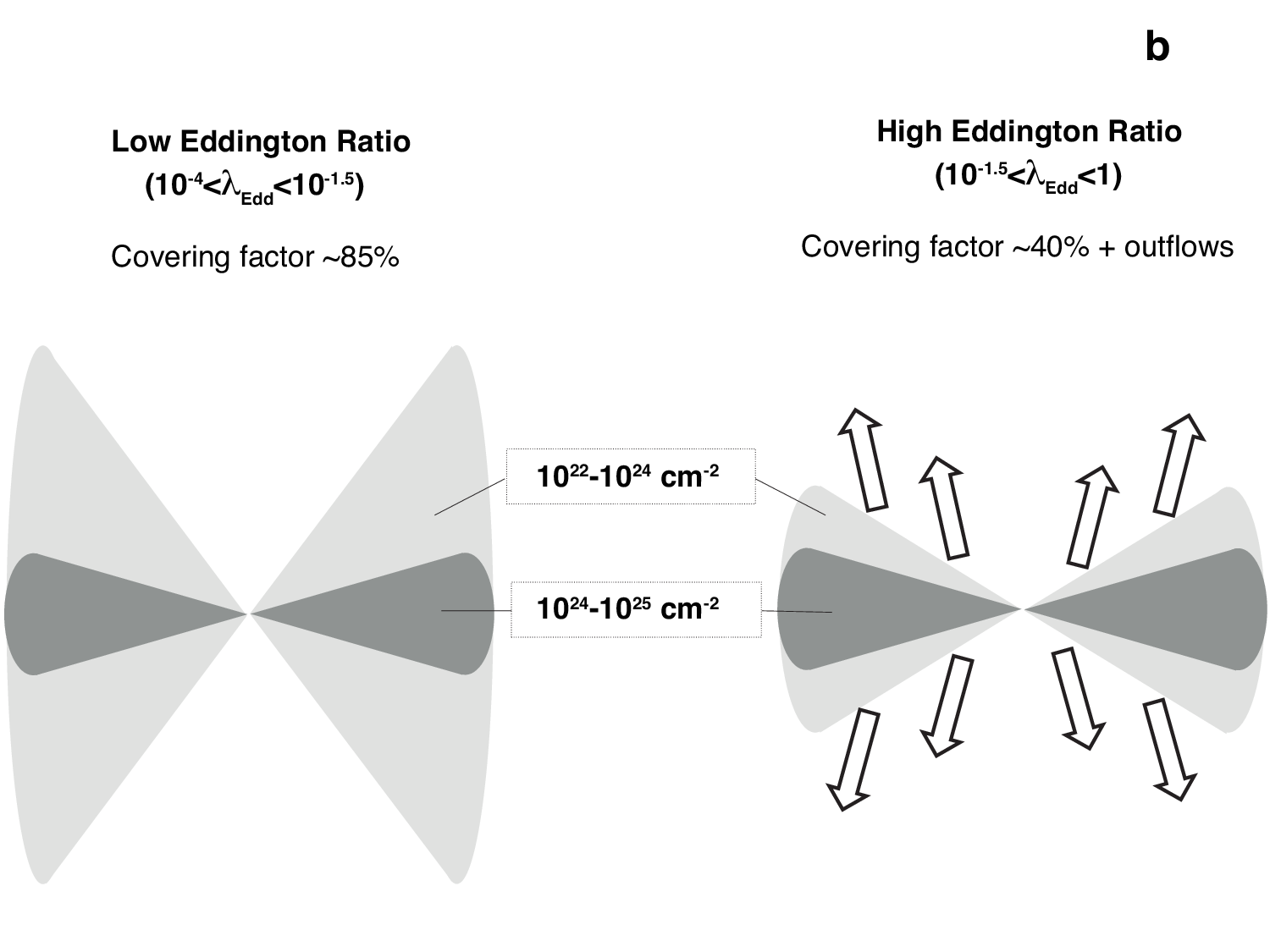}
 \caption{{\it Left panel}: relationship between the covering factor of dusty gas and the Eddington ratio for nearby AGNs. In the framework of the radiation-regulated AGN unification, applicable for $10^{-4} \leq \lambda_{\rm Edd} < 1$, the Eddington ratio serves as the primary parameter influencing the likelihood of a source being obscured. The black solid curve represents the covering factor for both Compton-thin (yellow region) and Compton-thick (green region) material, with the shaded regions denoting the 1-$\sigma$ uncertainties. {\it Right panel}: schematic diagram depicting the distribution of material surrounding supermassive black holes across different Eddington ratio ranges, in the framework of the radiation-regulated unification model. Figure from \cite{Ricci:2017rn}.}
 \label{fig:RRUM}
\end{figure}

These methods have also highlighted important correlations between the covering factor and AGN physical parameters. Early studies identified a relationship between the torus covering factor and AGN luminosity, as evidenced by the decreasing fraction of Type-2 AGNs with increasing luminosity \citep{Lawrence:1982bt,Lawrence:1991ah}. This trend has been confirmed using X-ray surveys \citep[e.g.,][]{Ueda:2003qf,Merloni:2014iw}, infrared observations \citep[e.g.,][]{Maiolino:2007ii,Ichikawa:2019dj}, and IR studies based on torus models \citep[e.g.,][]{Alonso-Herrero:2011tz}. More recent works suggest that the Eddington ratio is the primary driver of this correlation \citep{Ricci:2017rn,Ricci:2022ke,Ricci:2023iu}, and that when controlling for Eddington ratio the luminosity dependence disappears. These works found a significant drop in the fraction of obscured sources at $\log \lambda_{\rm Edd} \gtrsim -2$ (left panel of Figure\,\ref{fig:RRUM}), aligning with the predicted Eddington limit for dusty gas at $N_{\rm H} \simeq 10^{22}\rm cm^{-2}$ \citep[e.g.,][]{Fabian:2006lq,Venanzi:2020gx}. This confirmed the idea that radiative feedback plays a crucial role in shaping the environments close to SMBHs: as the Eddington ratio increases, radiation pressure on dusty gas reduces the covering factor. This forms the foundation of the {\it radiation-regulated unification model} proposed by \citet{Ricci:2017rn}. According to this scheme, AGNs with low Eddington ratios ($\log \lambda_{\rm Edd} \lesssim -1.5$) are more likely to appear obscured. At higher Eddington ratios ($\log \lambda_{\rm Edd} \gtrsim -1.5$), however, radiation pressure removes material near the AGN, opening more unobscured lines of sight to the accreting source (right panel of Figure\,\ref{fig:RRUM}). This was later confirmed by studies of the Eddington ratio distribution function, which found clear differences between obscured and unobscured AGN (\citealp{Ananna:2022fg}, left panel of Figure\,\ref{fig:Evolution}). This process not only clears dust from the obscurer but may also drive dusty outflows, potentially explaining the polar mid-IR emission observed in many AGNs (see \S\ref{sect:polar}; see also \citealp{Alonso-Herrero:2021yl}). These findings suggest that, in addition to the inclination angle, the likelihood of observing a source as obscured or Type\,2 also depends on the AGN's Eddington ratio: at low $\lambda_{\rm Edd}$ is easier to have obscured/Type\,2 lines of sight than at high $\lambda_{\rm Edd}$.

\section{Evolutionary models, mergers and redshift evolution}\label{sect:evolutionMergRed}

The accreting phase of SMBHs represents only a transient stage in their evolutionary lifetimes (e.g., \citealp{Shen:2007ab,Shankar:2009xo}). During this growth phase, the properties of the circumnuclear material are likely to evolve (e.g., \citealp{Hopkins:2006cm}), influenced by external factors as well as feedback from the accreting black hole (\S\ref{sect:RRunification}). This is particularly important for the evolution of AGN, as the circumnuclear gas serves as the fuel enabling SMBH growth. In this Section we describe some evolutionary aspects of unification, in which the properties of AGN and the host galaxy can affect the properties of the nuclear environment of accreting SMBHs, as well as their observed properties.

\subsection{Radiation-regulated SMBH growth}\label{sect:RadRegulatedGrowth}

Radiative feedback expels the fuel from accreting SMBHs at high Eddington ratios, which suggests that the radiation-regulated unification model could be dynamic, with AGNs evolving across the $N_{\rm H}-\lambda_{\rm Edd}$ diagram during their lifetimes (\citealp{Ricci:2017rn}; see also \citealp{Jun:2021rr}). \citet{Ricci:2022ke} showed that the breaks in both the luminosity function and the Eddington ratio distribution function of nearby AGN \citep{Ananna:2022fg} correspond to the Eddington ratios and X-ray luminosities where AGNs transition from being largely obscured by absorbing material to having most of their sky free of obscuration. This shows that, at least at $z \sim 0$, SMBH growth primarily occurs during phases when the AGN is heavily obscured by gas and dust, while AGNs accreting above the Eddington limit for dusty gas are relatively rare. The rapid decrease in the AGN number density as $\lambda_{\rm Edd}$ increases above the Eddington limit for dusty gas suggests that, while removing the obscuring material, radiation pressure also depletes the fueling reservoir, thus regulating SMBH accretion and reducing the lifetime of high Eddington ratio AGNs \citep{Ricci:2022ke} .
\smallskip

\begin{figure*}
\centering
 %% 1st image
 %% 2nd image
\includegraphics[angle=90,height=0.35\textwidth]{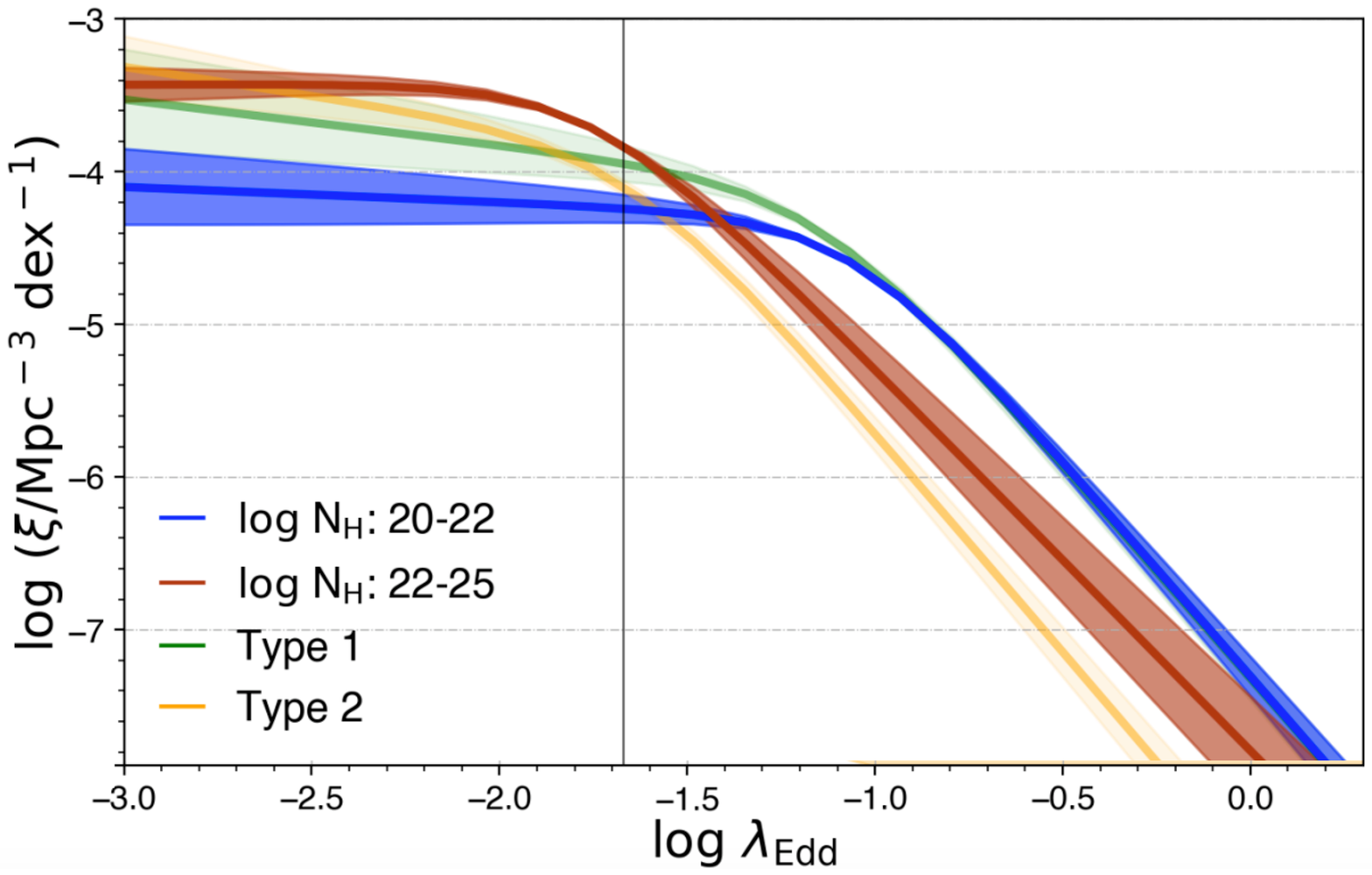}
\includegraphics[width=0.48\textwidth]{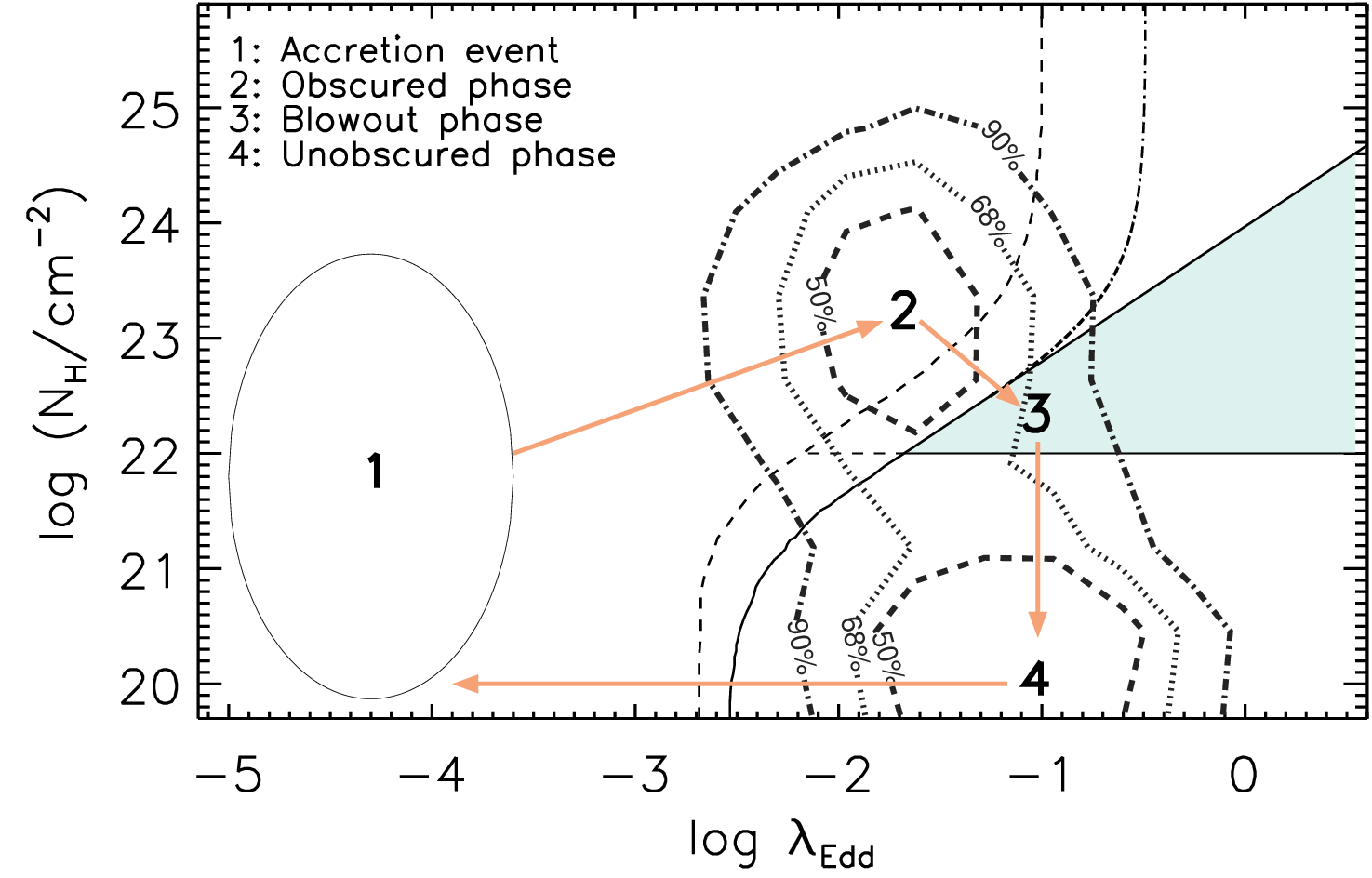}
% %% caption
% \begin{minipage}[t]{1\textwidth}
  \caption{{\it Left panel:} The Eddington ratio distribution functions for AGNs are shown for two column density bins: $\log(N_H/\mathrm{cm}^{-2}) = 20-22$ and $\log(N_H/\mathrm{cm}^{-2}) = 22-25$, as well as for Type 1 and Type 2 AGNs. The shaded regions represent the 1$\sigma$ uncertainty ranges. Figure from \cite{Ananna:2022fg}. {\it Right panel:} a schematic of the radiation-regulated growth model outlined here: an accretion event (1) leads to an increase of the Eddington ratio and typical column density of an AGN (2), which would be preferentially observed as an obscured or type\,2 source, due to the large covering factor of the obscuring material. As the Eddington ratio increases above the effective Eddington limit for dusty gas, the AGN spends a short time in the blowout region (3), before its covering factor decreases due to the effect of radiative feedback, and it is mostly observed as an unobscured or type\,1 source (4). Once most of the material has been accreted, or blown away by infrared radiation trapping, the source moves back to having low values of $N_{\rm H}$ and $\lambda_{\rm Edd}$. Figure from \citet{Ricci:2022ke}.}
\label{fig:Evolution}
% \end{minipage}
\end{figure*}

\noindent The differences in the Eddington ratio distributions between obscured and unobscured AGNs, along with the overall shapes of the Eddington Ratio Distribution Function and luminosity function (left panel of Figure\,\ref{fig:Evolution}), can be interpreted within the framework of an evolutionary model. In this scheme, radiation pressure shapes the SMBH's immediate environment and regulates its growth. The right panel of Figure\,\ref{fig:Evolution} present a schematic representation of this radiation-regulated growth model (\citealp{Ricci:2022ke}; see also \citealp{Ricci:2023iu}), an extension of the radiation-regulated unification model (\S\ref{sect:RRunification}). In this model:
\smallskip
\begin{enumerate}
    \item  SMBHs begin their growth at low $\lambda_{\rm Edd}$ in a predominantly unobscured state. An accretion event initiates an increase in Eddington ratio, column density, and covering factor of the circumnuclear obscuring material. The accretion event driving this cycle could result from either secular processes \citep[e.g.,][]{Davies:2007qo} or mergers \citep[e.g.,][]{Blecha:2018gt}, or a combination of both. 
\smallskip
    \item As material flows toward the SMBH, $\lambda_{\rm Edd}$, $N_{\rm H}$, and the covering factor increase. Due to the high covering factor, an AGN in this phase would typically appear obscured to an observer at a random inclination angle.
\smallskip
    \item As the Eddington ratio increases, the AGN reaches the effective Eddington limit for dusty gas and enters the blowout region. In this phase, the covering factor decreases rapidly as most of the circumnuclear obscuring material is blown away.
\smallskip
    \item Once most of the material is removed, the AGN transitions to a phase with lower column density and covering factor, during which is more likely to be observed as unobscured. This final stage is expected to be brief, lasting approximately $3-30$\,Myr \citep{Ananna:2022fg}, assuming an AGN lifetime of $\sim 10^7-10^8$\,yr.  
\end{enumerate}
\smallskip

\begin{figure*}
\centering
 %% 1st image
 %% 2nd image
\includegraphics[angle=90,height=0.65\textwidth]{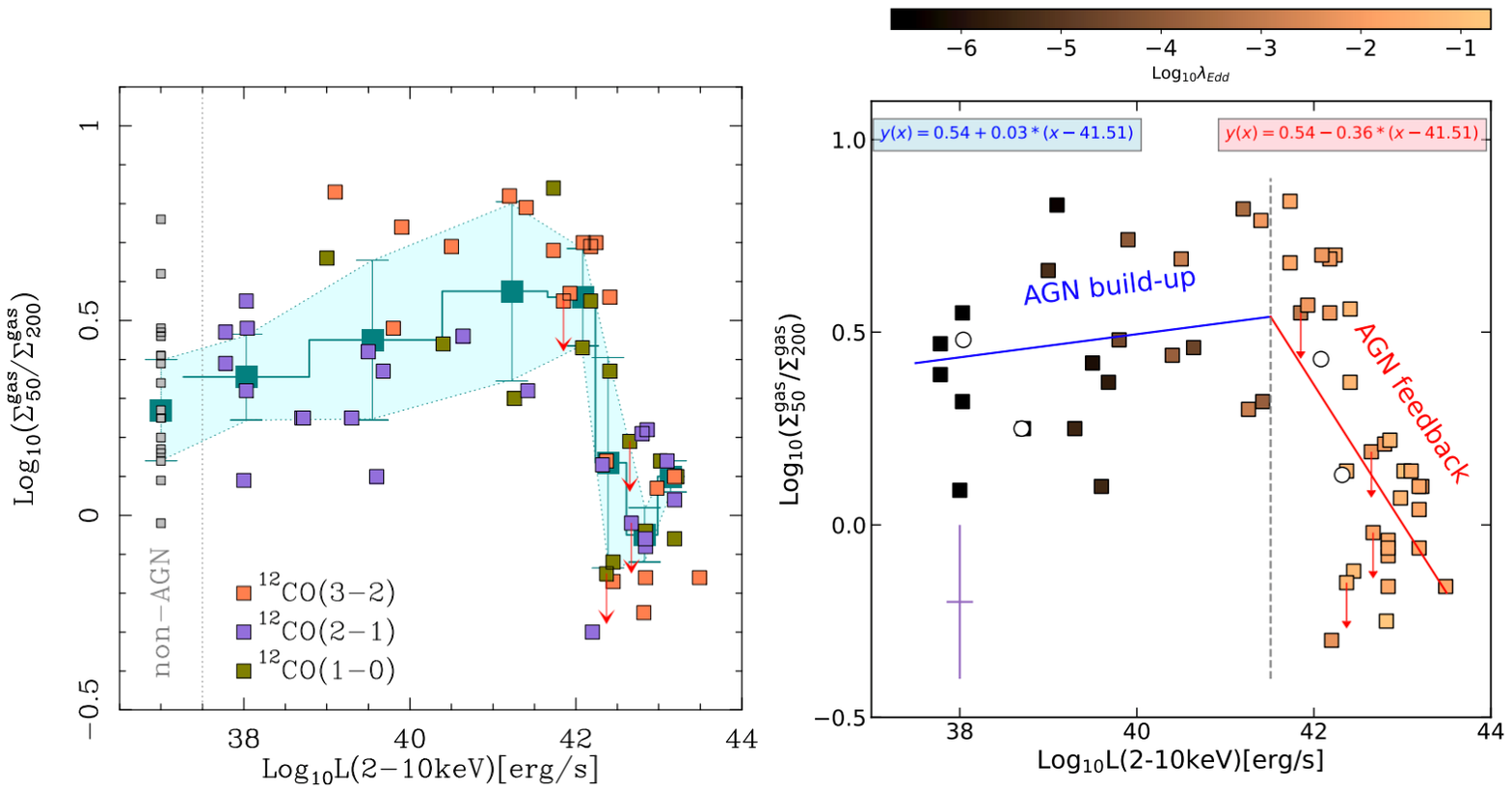}
% %% caption
\vspace{-1.5cm}
% \begin{minipage}[t]{1\textwidth}
  \caption{{\it Left panel:} Concentration of cold molecular gas at two spatial scales: $r \leq 50$ pc ($\Sigma_{\rm gas, 50}$) and $r \leq 200$ pc ($\Sigma_{\rm gas, 200}$), corresponding to the nuclear and circumnuclear disk regions. The molecular gas concentration is plotted as a function of AGN 2--10 keV  X-ray luminosities. Individual data points (small squares) are color-coded based on the CO transition used to estimate the surface densities for each AGN. For non-AGN targets, concentration indices are represented as gray squares positioned arbitrarily at $\log(L_X) = 37$ erg s$^{-1}$. The large green squares indicate the median concentration index. The green-shaded region shows the uncertainties.
\textit{Right panel:} Same as the left panel but showing the two-branch linear solution. The blue line represents the AGN build-up phase, while the red line corresponds to the AGN feedback phase. Symbols are color-coded based on the Eddington ratio, except for  galaxies lacking this information, which are shown as empty circles. Figures from \cite{Garcia-Burillo:2024vy}. }
\label{fig:Evolution2}
% \end{minipage}
\end{figure*}

\noindent In the final stage, the remaining obscuring material is expected to be optically thick and concentrated near the disk plane. This aligns with observations showing that the covering factor of Compton-thick gas remains relatively constant across different Eddington ratios (\citealp{Ricci:2017rn}; see Figure\,\ref{fig:RRUM}). Gradually, as the remaining gas and dust is accreted, or removed by radiation pressure and infrared trapping, the AGN shifts to a state characterized by low $N_{\rm H}$ and low $\lambda_{\rm Edd}$ (i.e., stage 1). In agreement with the idea that radiative feedback is important in nearby AGN, \citet{Garcia-Burillo:2021ix} and \citet{Garcia-Burillo:2024vy} recently analyzed ALMA observations of nearby AGNs and non-AGN galaxies and found a distinct turnover in cold molecular gas concentration at a breakpoint of $L_X \sim 10^{41.5} \, \mathrm{erg \, s^{-1}}$, separating two regimes: (1) the "AGN build-up branch" ($L_X \leq 10^{41.5} \, \mathrm{erg \, s^{-1}}$) with high gas concentrations and peaked radial profiles, and (2) the "AGN feedback branch" ($L_X \geq 10^{41.5} \, \mathrm{erg \, s^{-1}}$) with reduced gas concentrations and flatter or inverted profiles (Figure\,\ref{fig:Evolution2}). They interpreted these findings, along with evidence of molecular deficits on nuclear scales, as signatures of AGN feedback and a possible evolutionary sequence linked to increasing luminosities and Eddington ratios.

\subsection{Mergers and Starformation}\label{sect:Mergers}

\begin{figure}[b!]
\centering
 %% 1st image
 %% 2nd image
\includegraphics[angle=90,width=14cm]{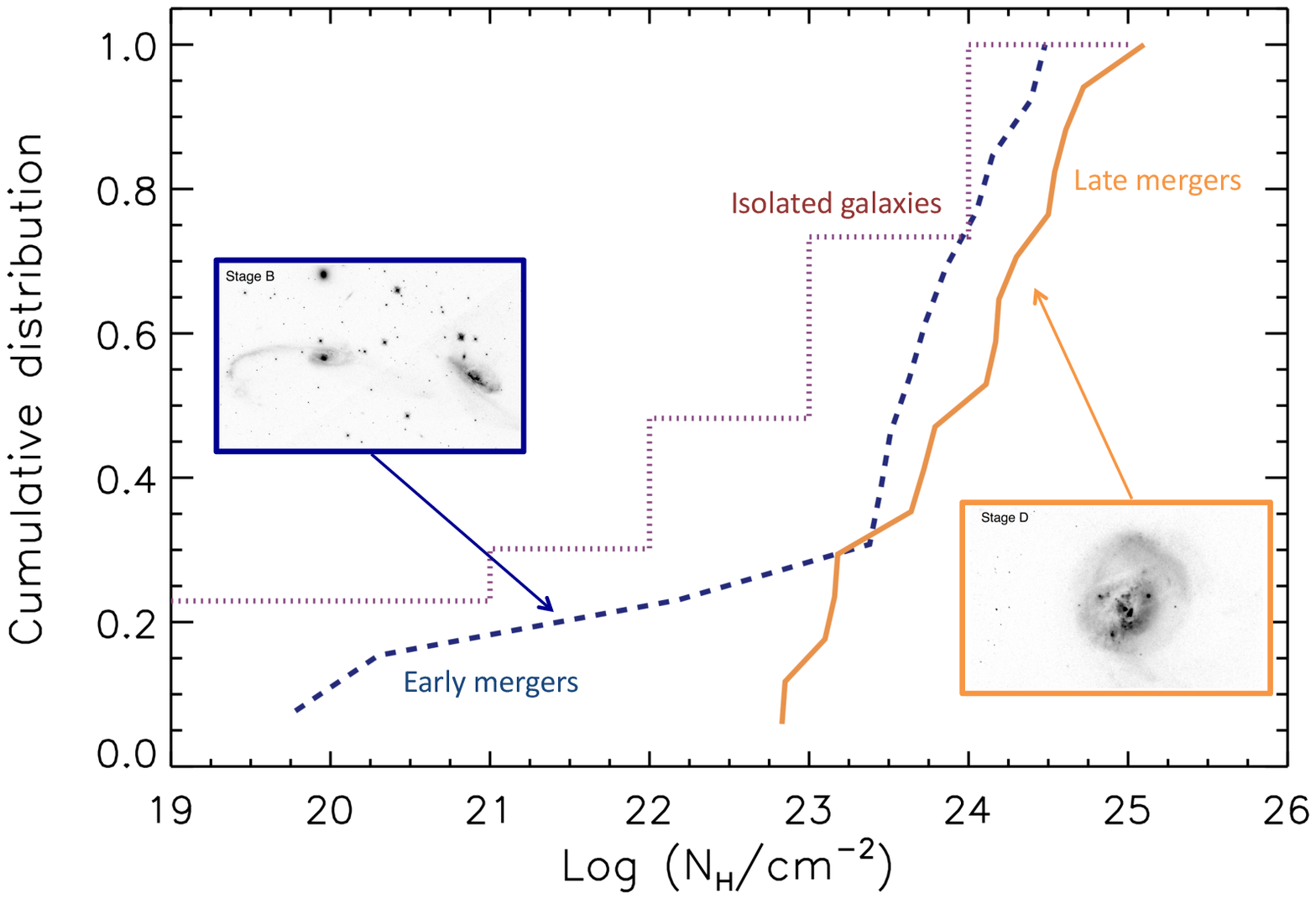}
% %% caption
  \caption{Cumulative distribution of column density for AGNs in early-stage mergers (dashed blue line) and in late-stage mergers (solid orange line). AGNs in the final stages of a merger show higher levels of obscuration compared to those in the early stages, as well as compared to nearby AGNs selected through hard X-rays (dotted purple line; \citealp{Ricci:2015fk,Ricci:2017aa}). Figure adapted from \cite{Ricci:2021oz}.}
\label{fig:NHmergers}
\end{figure}

Galaxy mergers could also play a crucial role in shaping the distribution of obscuring material around SMBHs. During the merging process, interactions between galaxies drive substantial inflows of gas toward the nucleus, increasing both obscuration and SMBH accretion (e.g., \citealp{De-Rosa:2019lm}). Simulations (e.g., \citealp{Blecha:2018gt}) suggest that column densities increase rapidly during mergers, with the most obscured phase occurring when the nuclei are at small separations. Supporting this, studies of local Luminous and Ultra-luminous Infrared Galaxies (U/LIRGs) in different merger stages have shown that AGNs in mergers are significantly more obscured than those in non-merging galaxies, with obscuration properties evolving throughout the merger \citep{Ricci:2017aa,Ricci:2021oz,Yamada:2021vz}. \citet{Kocevski:2015zr} found an increased incidence of merger activity and interaction features among AGNs with $N_{\rm H} > 3 \times 10^{23} \,\mathrm{cm^{-2}}$ at $z \sim 1$ compared to unobscured AGNs.
Broad-band X-ray observations reveal that the fraction of Compton-thick (CT) AGNs is much higher in late-stage mergers ($\sim 60-70\%$) compared to local hard X-ray selected AGNs ($\sim 30\%$, see Figure\,\ref{fig:NHdistribution}; \citealp{Ricci:2017aa,Ricci:2021oz}). AGN obscuration peaks when the nuclei of the merging galaxies are separated by projected distances of $\sim 0.4-10.8\,\mathrm{kpc}$, where the CT AGN fraction reaches $77^{+13}_{-17}\%$. The most extreme of these sources could be associated with Compact Obscured Nuclei (or CONs, \citealp{Aalto:2015fu}), which are thought to host AGNs fully embedded in gas with extreme column densities ($>10^{25}\rm\,cm^{-2}$). The high obscuration levels found in merging galaxies highlight a potential selection bias, as lower luminosity AGNs in advanced mergers may remain undetected due to extreme obscuration \citep{Ricci:2021oz}. Moreover, in a  study of galaxy pairs, \citet{Satyapal:2014oq} observed a higher fraction of IR-selected AGNs compared to optically-selected AGNs in advanced merger stages, suggesting that significant obscuration may cause many AGNs to be missed by optical classifications.
One key finding from recent studies is that nearly all AGNs in late-stage mergers have $N_{\rm H} > 10^{23}\,\mathrm{cm^{-2}}$, indicating that $95^{+4}_{-8}\%$ of the X-ray sources are heavily obscured \citep{Ricci:2017aa,Ricci:2021oz}. In contrast, the fraction of obscured sources is lower in galaxies at earlier merger stages, and both differ significantly from the obscuration properties observed in mostly non-merging, nearby AGNs (see Figure\,\ref{fig:NHmergers}). This suggests that the covering factor of the obscuring material in late-stage mergers is extremely high, effectively enveloping the AGN and obscuring almost all lines of sight to the accreting SMBH. Further evidence for high covering factors in advanced mergers comes from studies of the [O\,IV]\,25.89$\mu$m line. \citet{Yamada:2019ld,Yamada:2024tk} showed that the ratio of [O\,IV] to 12$\mu$m AGN luminosity decreases as mergers progress, implying that the covering factor of the obscuring material increases in late-stage mergers. All this strongly suggests that the host galaxy properties, and in particular its merging state, can be a fundamental parameter in determining the probability of an AGN to be observed as obscured/Type\,2.
\smallskip

\noindent In addition to mergers, gas inflows into the nuclear regions on sub-100-pc scales can lead to the formation of compact starburst disks, which serve as reservoirs for both SMBH growth and intense star formation. These starbursts can significantly contribute to AGN obscuration \citep{Ballantyne:2008ow,Andonie:2024tw}. Furthermore, obscuration has been found to correlate with the stellar mass of the host galaxy \citep{Buchner:2017jx}, suggesting a connection between the interstellar medium and the obscuration of the accreting SMBH.

\smallskip
\noindent Some of the obscuring gas in merging or strongly star-forming systems may reside at distances of hundreds of parsecs from the accreting SMBH and beyond its sphere of influence. This would make luminosity, rather than Eddington ratio, the relevant parameter in regulating radiation pressure effects. Moreover, stars embedded within the dusty gas clouds can provide additional gravitational binding, counteracting radiation pressure and allowing the obscuring material to persist at higher $\lambda_{\rm Edd}$ \citep{Fabian:2009ez}. Considering all this, in major merger-driven accretion scenarios, the radiation-regulated growth model (right panel of Figure\,\ref{fig:Evolution}) predicts that the AGN can reach higher column densities and accretion rates.

\begin{figure}
\centering
 %% 1st image
 %% 2nd image
\includegraphics[angle=90,width=0.48\textwidth]{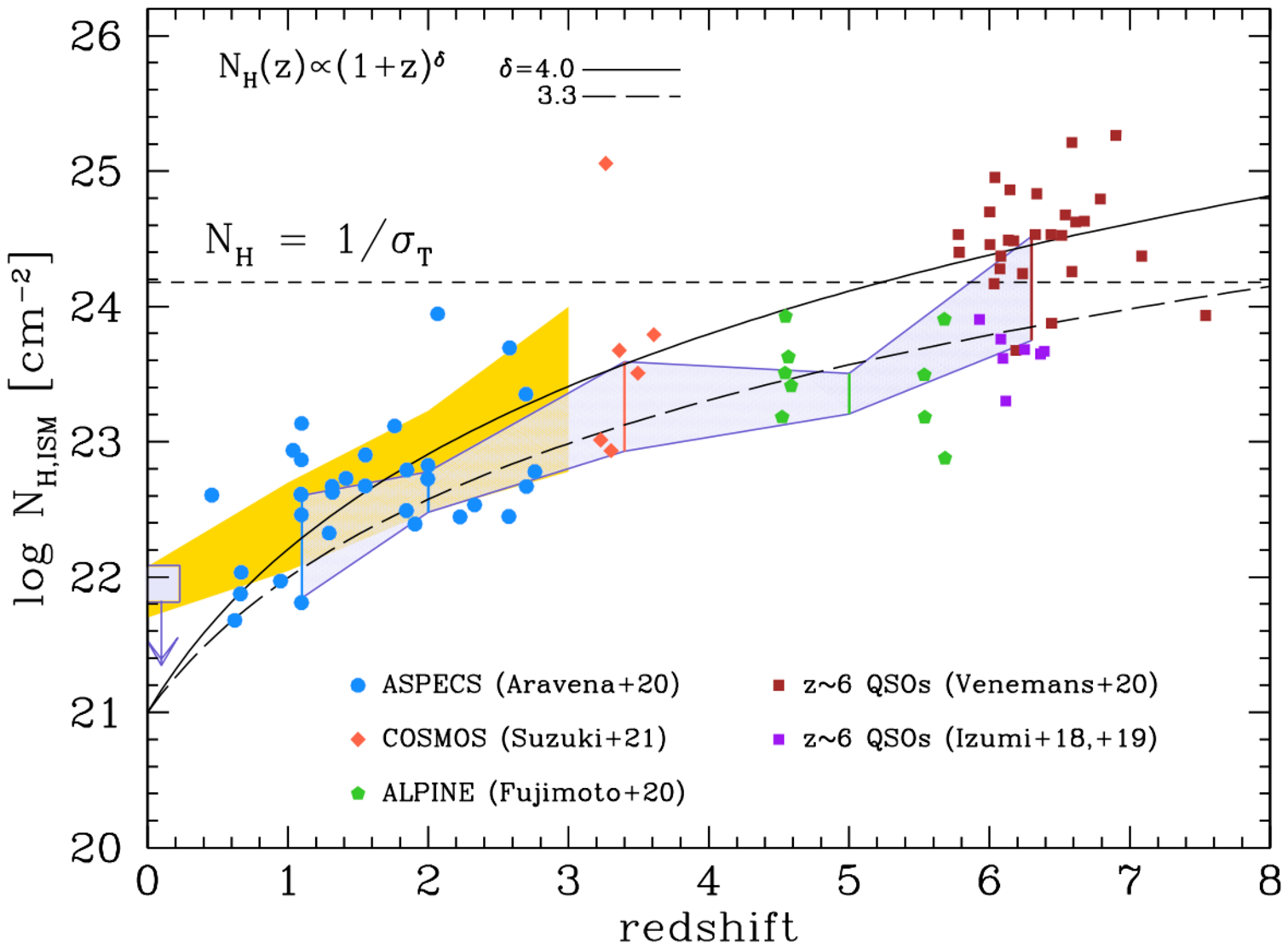}
\includegraphics[angle=90,width=0.48\textwidth]{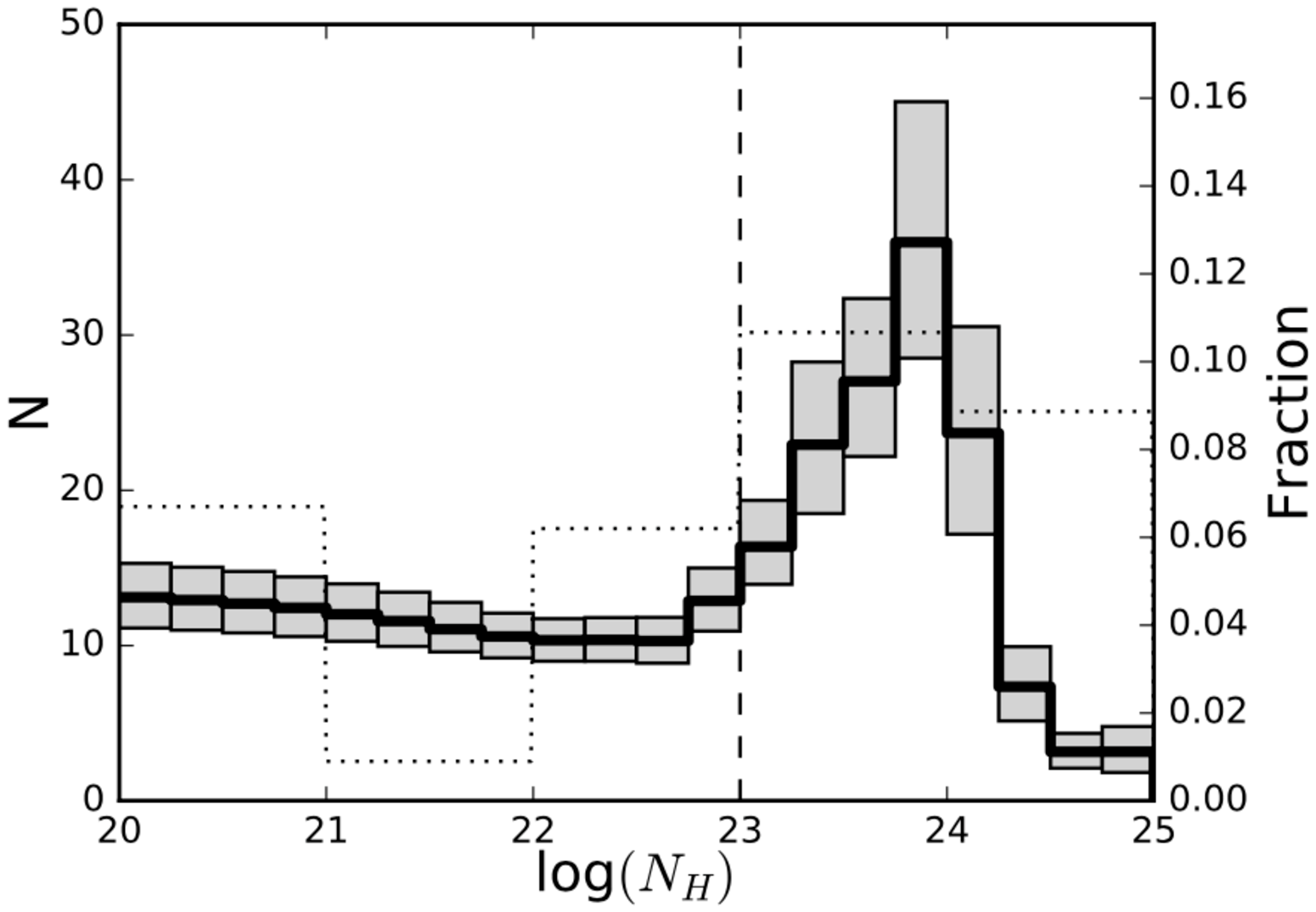}
% %% caption
  \caption{{\it Left panel:} Column density of the interstellar medium vs redshift for different samples of massive galaxies. Figure from \cite{Gilli:2022pu}. {\it Right panel:} Intrinsic column density distribution for AGN at $ 3 \leq z < 6$. Figure from \citet{Vito:2018ot}.}
\label{fig:highz}
\end{figure}

\subsection{Evolution with redshift}\label{sect:redshiftevolution}

As discussed in \S\ref{sect:RadRegulatedGrowth}, radiation pressure seems to play a dominant role in regulating nuclear obscuration in nearby AGN, and accreting SMBHs tend to avoid the blowout region in the $N_{\rm H}-\lambda_{\rm Edd}$ plane (cyan region in the right panel of Figure\,\ref{fig:Evolution}). However, this does not seem to necessarily apply at redshifts and luminosities higher than those discussed in \S\ref{sect:RadRegulatedGrowth} and \ref{sect:Mergers}, where sources accreting at very high Eddington ratios ($\lambda_{\rm Edd}$) can be frequently obscured. Two clear examples are given by Hot Dust Obscured Galaxies (Hot DOGs) and red quasars, typically found at $z\gtrsim 1$, which have been found to be in the "blowout" region of the $N_{\rm H}$--$\lambda_{\rm Edd}$ diagram (e.g., \citealp{Glikman:2017ug}). These objects have also been found to show strong outflows, suggesting that material is actively being expelled from the nuclear regions \citep{Temple:2019pr}. Similarly, a recent study by \citet{Jun:2021rr}, examining a large sample of infrared and submillimeter-bright obscured quasars with bolometric luminosities $ \gtrsim 10^{46}\,\mathrm{erg\,s^{-1}}$ found that most of these objects reside in the blowout region. These systems might represent a short-lived phase where the AGN is expelling dusty circumnuclear gas and clearing its immediate environment. The behavior of these AGN may also be linked to the high fraction of mergers identified in these systems \citep[e.g.,][]{Urrutia:2008vn}, which can enhance both SMBH accretion and obscuration (see \ref{sect:Mergers}).
\smallskip

\noindent At even higher redshifts, the interstellar medium may also play a significant role in obscuring AGNs. \citet{Gilli:2022pu} used ALMA observations to show that ISM column densities toward nuclei in $z > 3$ galaxies are typically more than one hundred times greater than those at $z \sim 0$, reaching Compton-thick levels at $z \gtrsim 6$ (left panel of Figure\,\ref{fig:highz}). Combined with the increased frequency of mergers at higher redshifts relative to $z \sim 0$ \citep[e.g.,][]{Mortlock:2013is}, this could explain the higher fraction of obscured AGNs observed at high redshift \citep[e.g.,][]{Ueda:2014ix,Buchner:2015ve}. In particular, a significant fraction of AGNs at $z > 3$ have very high column densities ($N_{\rm H} \gtrsim 10^{23}\,\mathrm{cm}^{-2}$; \citealp{Vito:2018ot}; left panel of Figure\,\ref{fig:highz}; see also \citealp{Peca:2023og}). Since the interstellar medium resides beyond the SMBH's sphere of influence, the relationship between radiative AGN feedback and obscuration at high redshifts would depend primarily on luminosity rather than Eddington ratio. This allows AGNs to reach higher luminosities before clearing the obscuring gas.  If part of the obscuring material is associated with AGN fueling, one would expect the break in the Eddington ratio distribution function to shift to higher values at higher redshifts, similar to the observed shift in the break of the AGN luminosity function \citep[e.g.,][]{Ueda:2014ix}. This would lead to an evolution with redshift of the radiation-regulated growth model discussed in \S\ref{sect:RadRegulatedGrowth}.

\bigskip

\noindent Based on the discussion above, it appears that AGN unification should account not only for inclination angle (\S\ref{sect:AGNclassification}) and Eddington ratio (\S\ref{sect:RRunification}) but also for redshift and host galaxy properties (e.g., merger stage, star formation rate, etc.; \S\ref{sect:Mergers} and \ref{sect:redshiftevolution}), as these factors can influence the probability of a source being obscured (see Figure\,\ref{fig:UpdatedUM}).

\section{Changing-look AGN}\label{sect:changinglook}

Another clear exception to the traditional orientation-dependent unification model is presented by Changing-Look AGN (CL AGN; \citealp{Matt:2003sw}, see \citealp{Ricci:2023fw} for a recent review). These sources exhibit dramatic and rapid spectral changes across different frequencies, challenging the static nature of the unification model. CL AGN are characterized by transitions between Type\,1/unobscured and Type\,2/obscured classifications within observable timescales. Such transitions manifest as significant changes in optical/UV emission lines or X-ray absorption properties. CL AGN can be divided into two main classes:
\smallskip

\noindent{\bf Changing-obscuration AGN (CO-AGN)}: primarily detected in the X-rays, these objects show variations in the line-of-sight column density (left panel of Figure\,\ref{fig:schematicCLAGN_spec}).
\smallskip

\noindent{\bf Changing-state AGN (CS-AGN):} primarily identified in the optical/UV, these AGN show the appearance or disappearance of broad emission lines (right panel of Figure\,\ref{fig:schematicCLAGN_spec}).
\smallskip

\noindent In the following we briefly summarize some of the main features of these two classes of objects.

\subsection{Changing-obscuration AGN (CO-AGN)}\label{sect:COAGN}
CO-AGN undergo transitions in their X-ray properties due to changes in the column density of obscuring material along the line of sight. These transitions, which can bring an AGN from a Compton-thin state to a Compton-thick state, are interpreted as evidence for clumpy and dynamic obscuration near the SMBH (e.g., \citealp{Torres-Alba:2023iz}). The variability in column density observed so far occurs on timescales from hours to years, suggesting that the obscuring material resides in regions associated with the BLR or the inner edge of the torus. The CO events can be associated with several mechanisms:
\smallskip

\begin{itemize}
    \item \textbf{Eclipses:}  Eclipses of the X-ray source are considered one of the primary causes of CO events, with variations in $N_{\rm H}$ are attributed to the movement of individual clouds within the BLR or torus \citep[e.g.,][]{Risaliti:2002mc,Matt:2003sw}. X-ray monitoring campaigns of nearby AGN, such as NGC\,1365, have detected these events on very short timescales, sometimes lasting only a few hours \citep{Risaliti:2009mi}, and with column densities varying up to $\sim 10^{24}\rm\,cm^{-2}$. Observations of these eclipses provide valuable insights into the location of the obscuring clouds responsible for the variability and allow for estimates of the size of the X-ray corona (e.g., \citealp{Risaliti:2007aa}).  
    \item \textbf{Extreme flux Variability:} In some cases, the dramatic changes in the X-ray spectral properties of AGN may be directly linked to variations in their intrinsic X-ray luminosity. A rapid decrease in accretion power would result in a decline in the primary X-ray continuum (the power-law component). However, reprocessed X-ray radiation originating from material on scales of approximately 0.1--1\,pc would respond more slowly to changes in the illuminating X-ray flux. During periods of low primary flux, this delayed response could cause the reprocessed emission to dominate, leading the AGN to appear as reflection-dominated and potentially be misclassified as Compton-thick (e.g., \citealp{Gilli:2000bw}). A subsequent increase in accretion-powered radiation could restore the AGN's continuum-dominated appearance. This switch-off/switch-on behavior might result in the AGN being misidentified as a CO-AGN.
    \item \textbf{Outflows:}  While most CO events involve variations in neutral absorption, some AGN, particularly unobscured/Type\,1 sources, exhibit variability caused by ionized absorbing gas along the line of sight. These absorbers are often linked to outflows, which are common in AGN and have been detected in both X-ray and optical/UV spectra \citep{King:2015bi}. For example, a 2013 multi-wavelength campaign on the Type\,1 AGN NGC\,5548 revealed a CO event with two obscuring components: a low-ionization absorber and a nearly neutral absorber. UV spectroscopy identified broad absorption lines from the low-ionization component, showing outflow velocities up to $\sim 5000\,\mathrm{km\,s^{-1}}$. 
\end{itemize}
\smallskip

\noindent CO-AGN are extremely useful to improve our understanding of the AGN structure, as they can provide important insights in the size of the X-ray corona, on the interplay between disk and outflows, as well as on the physical and kinematical properties of the obscuring clouds.

\begin{figure*}
\centering
\includegraphics[angle=90,width=0.68\textwidth]{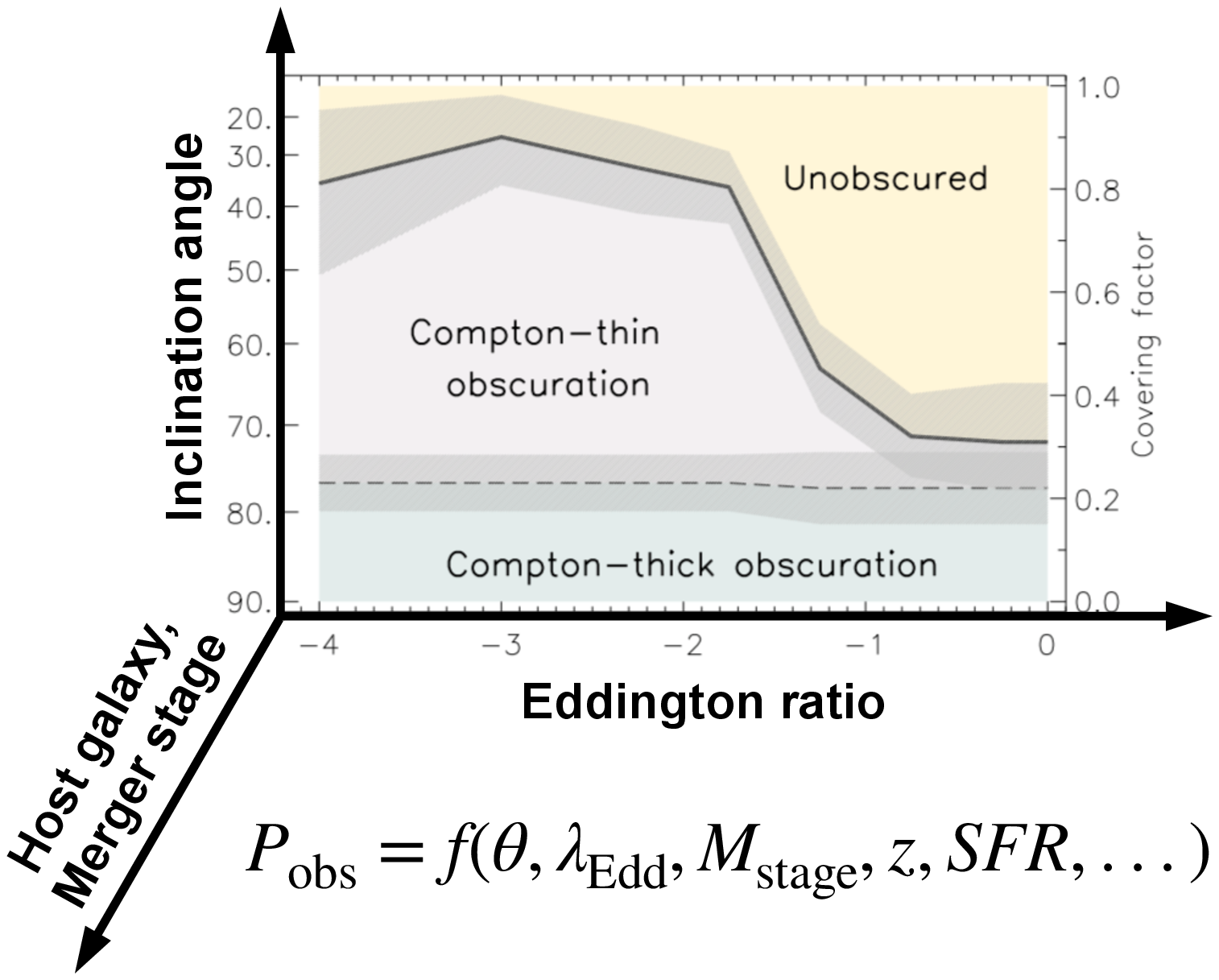}
\caption{A revised version of the unification model, incorporating Eddington ratio and host galaxy properties (e.g., merger stage, star formation rate, etc.) on the probability of a source being obscured. Adapted from \cite{Ricci:2017rn}.}
\label{fig:UpdatedUM}
\end{figure*}

\subsection{Changing-state AGN (CS-AGN)}\label{sect:CSAGN}

\begin{figure*}
\centering
\includegraphics[angle=90,width=0.68\textwidth]{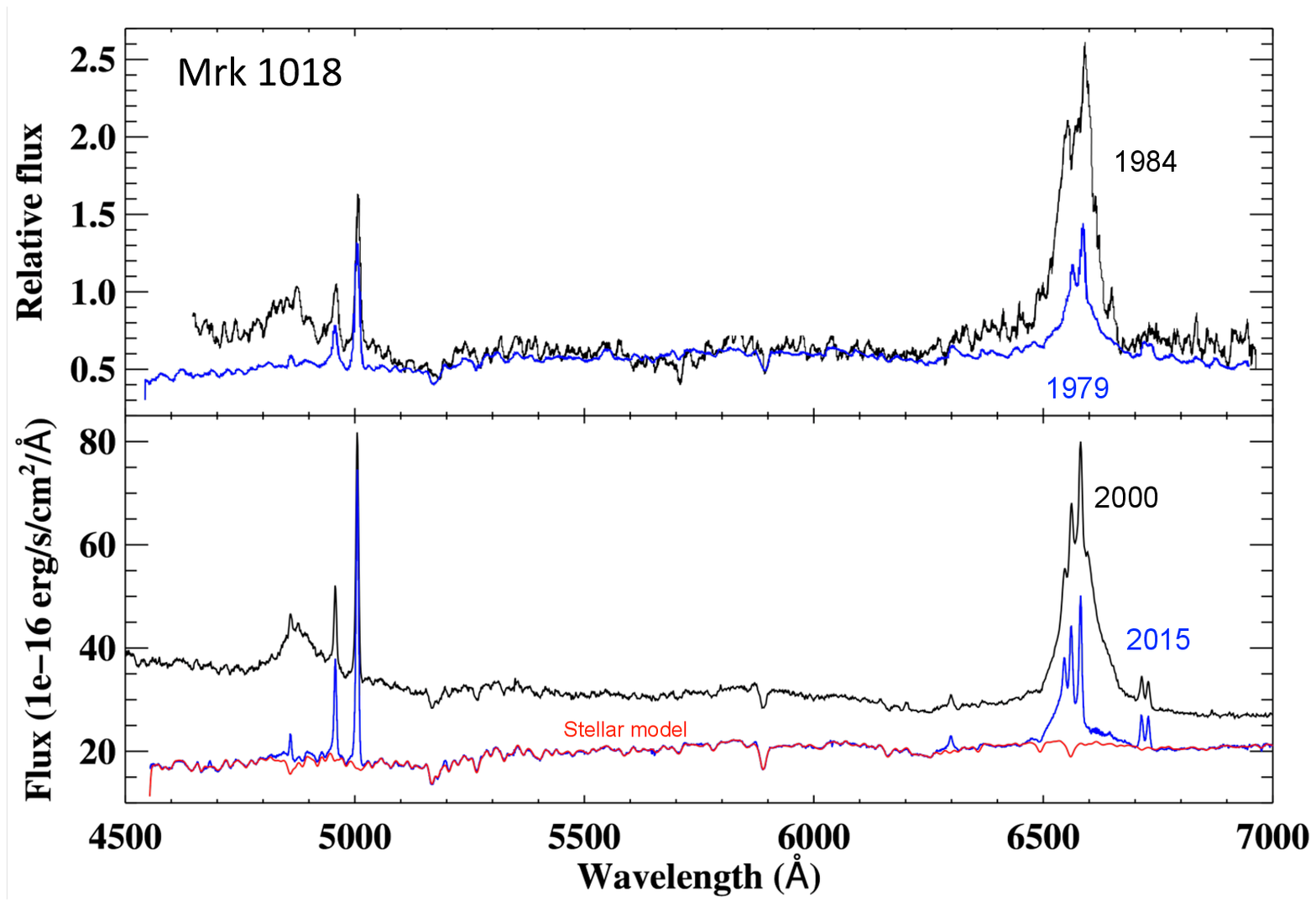}
\caption{Repeated spectral variations in Mrk\,1018 observed from 1979 to 2015. Figure from \cite{McElroy:2016hs}.}
\label{fig:Mrk1018}
\end{figure*}

CS-AGN undergo transitions between Type\,1 and Type\,2, driven by intrinsic changes in accretion rates. These transitions manifest as the appearance or disappearance of broad emission lines and significant optical/UV flux variability (e.g., \citealp{Tohline:1976gd,Husemann:2016iu,Komossa:2026ro}). A few CS-AGN have been observed to complete a full cycle, transitioning twice between classifications. For example, Mrk\,1018 transitioned from Type\,1.9 to Type\,1 in less than five years \citep{Cohen:1986sa} and reverted back to Type\,1.9 after 30 years (\citealp{McElroy:2016hs}, see Figure\,\ref{fig:Mrk1018}). In recent years, extensive imaging and spectroscopic surveys have significantly increased the number of identified CS-AGN, broadening the range of redshifts and key SMBH properties explored (e.g., \citealp{Green:2022cv,Zeltyn:2024zm}). Some of these surveys have discovered peculiar objects, such as 1ES\,1927+654 \citep{Trakhtenbrot:2019ay}, with properties very different to typical AGN \citep{Ricci:2020fp}.
CS transitions have been shown to generally not result from varying obscuration (see \citealp{Ricci:2023fw} for a review on the topic) but are more likely triggered by changes in the accretion rate associated to one of two mechanisms:

\smallskip

\begin{itemize}
    \item \textbf{Disk Instabilities:}  Disk instabilities could provide an explanation for CS-AGN transitions, particularly the connection between broad-line disappearance and soft X-ray excess variability. Observations of Mrk\,1018 reveal that such transitions are associated with a significant decrease in the soft X-ray excess, while harder X-ray flux remains relatively stable \citep{Noda:2018jz}. This spectral hardening resembles soft-to-hard state transitions in X-ray binaries, driven by disk instabilities near $\lambda_{\rm Edd} \simeq 10^{-2}$. Similar behaviors have been observed in other CS-AGN (e.g., \citealp{MacLeod:2019rk,Temple:2023sp,Jana:2024sb,Jana:2025wq,Jana:2026hx}), possibly confirming the role of spectral transitions between thin disks and ADAF-like flows \citep{Noda:2018jz}.
    \item \textbf{Major disk perturbations:}  Tidal disruption events (TDEs) are a compelling explanation for some CS-AGN, as illustrated by SDSS\,J0159+0033, where a decline in flux following an outburst matched the $t^{-5/3}$ trend expected for TDEs \citep{Merloni:2015ew}. In 1ES\,1927+654, TDE-like interactions may explain the observed disappearance and reappearance of the X-ray corona and peculiar broad-line behavior, highlighting how such events can significantly disrupt the accretion disk \citep{Ricci:2020fp,Li:2022pp}, leading to super-Eddington accretion episodes \citep{Li:2024zh}. Other mechanisms, such as interactions in binary SMBH systems or perturbations from stellar-mass objects, have also been proposed as potential triggers for CS events, though their contributions require further exploration (e.g., \citealp{Wang:2020si}).
\end{itemize}

\smallskip

\noindent CS transitions present a timescale problem, as the viscous timescale of accretion disks, which is typically expected to govern such changes, is considerably longer than the observed timescales of CS events. This discrepancy may be explained by factors such as an increased disk scale height or a higher sound speed in AGN disks due to radiation pressure, both of which can reduce the viscous timescale \citep{Noda:2018jz}. Additionally, magnetic pressure may further shorten these timescales. CS-AGN can be extremely important for improving our understanding of AGN accretion flows and their relationship with the surrounding circumnuclear material. They allow us to study phenomena such as the creation and evolution of the corona and BLR, as well as instabilities and perturbations within the accretion flow.

\section{Summary and future prospects}

\noindent In this chapter, we have provided a review of AGN unification models. We discussed the fundamental components of accreting SMBHs (\S\ref{sect:componennts}), including the accretion disk, broad-line and narrow-line regions, jets, outflows, and the dusty anisotropic obscurer. Traditional orientation-based unification models were revisited (\S\ref{sect:AGNclassification}) alongside more recent advancements, which have highlighted the important roles of covering factor and Eddington ratio in shaping the observational properties of AGN (\S\ref{sect:RRunification}). The role of radiative feedback might imply that nuclear obscuration changes over time, and that AGN move around in the column density-Eddington ratio diagram during their lives (\S\ref{sect:evolutionMergRed} and Fig.\,\ref{fig:Evolution}). We have also discussed changing-look AGN and their crucial role to better understand SMBH accretion and the properties of the circumnuclear material (\S\ref{sect:changinglook}). Considering recent findings, we have highlighted how AGN unification should account not only for inclination angle and Eddington ratio but also for redshift and host galaxy properties such as merger stage, as these factors can influence the probability of a source being obscured (see Figure\,\ref{fig:UpdatedUM}; \S\ref{sect:evolutionMergRed}).
\smallskip

\noindent While significant progress in our understanding of AGN unification and obscuration has been made over the past decade, several challenges remain. The complex relationship between AGN obscuration and host galaxy properties, including the roles of mergers and starburst activity, is yet to be fully understood. Similarly, we still lack a clear understanding of the impact of a redshift evolution, with possibly elevated interstellar medium densities and higher merger rates, on AGN obscuration and evolution at high redshifts. Finally, shedding light on the physical mechanisms and typical timescales governing changing-state transitions will be fundamental, as these phenomena could provide us with crucial insights into the physical mechanisms at play in AGN accretion and variability.
\smallskip

\noindent In the future, facilities such as for example the Vera Rubin Observatory, \textit{JWST}, {\it ULTRASAT}, {\it eROSITA}, the Extremely Large Telescope, the {\it Einstein probe}, the {\it Roman Space Telescope} and \textit{NewAthena} will provide unprecedented time-domain, multi-wavelength, and high-resolution data on nearby and distant accreting SMBHs, enabling detailed studies of AGN variability, obscuration, and feedback processes.

\bibliographystyle{Harvard}
\bibliography{reference}

\end{document}